\documentclass[11pt]{elsarticle}
\usepackage{verbatim,amsmath,amssymb}
\usepackage{epsfig,float,color}
\usepackage{geometry}
\usepackage{setspace}
\usepackage{natbib}
\usepackage{amsthm,mathrsfs,amsfonts,dsfont}
\usepackage{hyperref}
\usepackage[yyyymmdd,hhmmss]{datetime}
\usepackage{cleveref}
\usepackage{caption}
\usepackage{subcaption}
\usepackage{tablefootnote}

\usepackage{placeins}
\usepackage{mmap}
\definecolor{darkblue}{rgb}{0,0,1}
\hypersetup{pdftex=true, colorlinks=true, breaklinks=true, linkcolor=darkblue, menucolor=darkblue, pagecolor=darkblue, citecolor=darkblue, urlcolor=darkblue}

\newcommand{\bitm}{\begin{itemize}}
\newcommand{\eitm}{\end{itemize}}
\newcommand{\bnumr}{\begin{enumerate}}
\newcommand{\enumr}{\end{enumerate}}

\newcommand {\sigab}{\sigma^{\alpha\beta}}

\newcommand {\aab}{a^{\alpha\beta}}
\newcommand {\auab}{a_{\alpha\beta}}

\newcommand {\augd}{a_{\gamma\delta}}

\newcommand {\Aab}{A^{\alpha\beta}}
\newcommand {\Auab}{A_{\alpha\beta}}

\newcommand {\tauab}{\tau^{\alpha\beta}}

\newcommand {\eqb}[1]{\begin{equation}\begin{array}{#1}}
\newcommand {\eqe}{\end{array}\end{equation}}

\newcommand {\esb}[1]{\begin{equation*}\begin{array}{#1}}
\newcommand {\ese}{\end{array}\end{equation*}}
\newcommand {\ds}{\displaystyle}

\newcommand {\pa}[2]{\frac{\partial{#1}}{\partial{#2}}}

\newcommand {\paqq}[3]{\frac{\partial^2{#1}}{\partial{#2}\,\partial{#3}}}
\newcommand {\back}{\! \! \!}
\newcommand {\is}{\back &=& \back}
\newcommand {\dis}{\back &:=& \back}

\newcommand {\plus}{\back &+& \back}
\newcommand {\mi}{\back &-& \back}

\newcommand {\tr}{\mathrm{tr}\,}

\newcommand {\dif}{\mathrm{d}}

\newcommand {\II}{{I\kern-.3em I}}
\newcommand {\III}{{I\kern-.3em I\kern-.3em I}}

\newcommand {\ba}{\boldsymbol{a}}

\newcommand {\bn}{\boldsymbol{n}}

\newcommand {\bx}{\boldsymbol{x}}
\newcommand {\by}{\boldsymbol{y}}

\newcommand {\bA}{\boldsymbol{A}}

\newcommand {\bC}{\boldsymbol{C}}

\newcommand {\bE}{\boldsymbol{E}}
\newcommand {\bF}{\boldsymbol{F}}

\newcommand {\bI}{\boldsymbol{I}}

\newcommand {\bM}{\boldsymbol{M}}
\newcommand {\bN}{\boldsymbol{N}}

\newcommand {\bR}{\boldsymbol{R}}
\newcommand {\bS}{\boldsymbol{S}}

\newcommand {\bU}{\boldsymbol{U}}

\newcommand {\bX}{\boldsymbol{X}}
\newcommand {\bY}{\boldsymbol{Y}}

\newcommand {\bsig}{\mbox{\boldmath$\sigma$}}

\newcommand {\bbC}{\mathbb{C}}

\newcommand {\bbH}{\mathbb{H}}

\newcommand {\IR}{{\rm\kern.24em
   \vrule width.02em height1.53ex depth-.05ex
   \kern-.3em R}}
\newcommand {\ic}{{\rm\kern.20em
   \vrule width.02em height1.0ex depth-.05ex
   \kern-.22em c}}
\newcommand {\ia}{{\rm\kern.20em
   \vrule width.02em height1.05ex depth-.0ex
   \kern-.25em a}}
\newcommand {\IC}{{\rm\kern.24em
   \vrule width.02em height1.4ex depth-.05ex
   \kern-.26em C}}
\newcommand {\ID}{{\rm\kern.34em
   \vrule width.02em height1.5ex depth-.05ex
   \kern-.36em D}}
\newcommand {\IS}{{\rm\kern.24em
   \vrule width.02em height1.6ex depth.05ex
   \kern-.26em S}}
\newcommand {\IT}{{\rm\kern.50em
   \vrule width.02em height1.55ex depth-.05ex
   \kern-.52em T}}

\newcommand {\IE}{{\rm\kern.24em
   \vrule width.02em height1.55ex depth-.05ex
   \kern-.33em E}}
\newcommand {\IEa}{{\rm\kern.24em
   \vrule width.02em height1.55ex depth-.05ex
   \kern-.33em E}^{1}_{ijkl}}
\newcommand {\IEb}{{\rm\kern.24em
   \vrule width.02em height1.55ex depth-.05ex
   \kern-.33em E}^{2}_{ijkl}}

\newcommand {\sI}{\mathcal{I}}
\newcommand {\sJ}{\mathcal{J}}

\newcommand {\sO}{\mathcal{O}}

\newcommand {\sS}{\mathcal{S}}

\newcommand {\Ass}[2]{\kern 0.9ex \vrule width0.45em height0.2ex depth0ex \kern -2.1ex \bigwedge_{#1}^{#2}}
\newcommand {\ASS}[2]{\kern 1.45ex \vrule width0.5em height0.2ex depth0ex \kern -2.65ex \bigwedge_{#1}^{#2}}

\graphicspath{ {/} }

\begin{document}

\begin{center}
\Large{\bf{A nonlinear hyperelasticity model for single layer blue phosphorus based on ab-initio calculations}}\\
\end{center}

\begin{center}

\large{Reza Ghaffari$^\S$\footnote{Email: ghaffari@aices.rwth-aachen.de},
       Farzad Shirazian$^\S$\footnote{Email: shirazian@aices.rwth-aachen.de},
       Ming Hu$^\dag$\footnote{Email: hu@sc.edu} and
       Roger A. Sauer$^\S$\footnote{Corresponding author, email: sauer@aices.rwth-aachen.de}}\\
\vspace{4mm}

\small{\textit{
$^\S$Aachen Institute for Advanced Study in Computational Engineering Science (AICES), \\
RWTH Aachen University, Templergraben 55, 52056 Aachen, Germany \\[1.1mm]
$^\dag$Department of Mechanical Engineering, University of South Carolina,\\
541 Main Street, Columbia, SC 29208, USA }}




\end{center}

\vspace{3mm}


\rule{\linewidth}{.15mm}
{\bf Abstract}: A new hyperelastic membrane material model is proposed for single layer blue phosphorus ($\beta$-P), also known as blue phosphorene. The model is fully nonlinear and captures the anisotropy of $\beta\text{-P}$ at large strains. The material model is calibrated from density functional theory (DFT) calculations considering a set of elementary deformation states. Those are pure dilatation and uniaxial stretching along the armchair and zigzag directions. The material model is compared and validated with additional DFT results and existing DFT results from the literature, and the comparison shows good agreement. The new material model can be directly used within computational shell formulations that are for example based on rotation-free isogeometric finite elements. This is demonstrated by simulations of the indentation and vibration of single layer blue phosphorus sheets at micrometer scales. The elasticity constants at small deformations are also reported.


\vspace{-4mm}
\rule{\linewidth}{.15mm}
\section{Introduction}
Two-dimensional materials are a fascinating group of materials that can have very different mechanical, electronic, chemical, and optical properties than their bulk form \citep{Mas-Balleste2011_01}. Since the discovery of graphene, scientists have looked for other interesting 2D materials, and phosphorus allotropes are very promising candidates due to their unique properties. Up to now, many stable two dimensional structures have been predicted for phosphorus \citep{Guan2014_01,Wu2015_01}. Single layer black phosphorus (also know as phosphorene), as the first of this family to be discovered, has a large range of applications in field effect transistors \citep{Li2015_01}, optoelectronic devices \citep{Guo2015_01}, sensors \citep{Kou2014_01}, batteries \citep{Li2017_01}, and energy storage \citep{Bagheri2016_01,Khandelwal2017_01}.  \citet{Zhu2014_01} predict the structure of single layer blue phosphorus. They use \textit{ab-initio} calculations to investigate its structure and predict a higher band gap for single layer blue phosphorus than for black phosphorus. For simplicity, black and blue phosphorus refer to two dimensional structures of phosphorus in the rest of this paper. \citet{Zhang2016_01} grow a single layer of blue phosphorus through epitaxial growth on Au(111) by using black phosphorous as precursor. While black phosphorus has a band gap of 0.3-0.4 eV, blue phosphorus has a much higher band gap of 2-3 eV \citep{Zhu2014_01,Xiao2015_01}, indicating its potential application in field effect transistors and optoelectronic devices. \\
\citet{Mogulkoc2018_01} study the electronic and optical properties of phosphide/blue phosphorus heterostructures and suggest their applicability in new generation optoelectric devices duo to their transparency to visible light and great absorption over the UV range. \citet{Sun2016_01} use density functional theory (DFT) to study thermomechanical properties of black and blue phosphorus. \citet{Li2015_01} study the application of single-layer and double-layer black and blue phosphorus in Li-Ion batteries and predict both of them can be good electrode materials due to their small diffusion energy barriers and ability to maintain their layered structures during lithiation and delithiation processes. \citet{Liu2015_01} investigate the effect of external strain on the electronic properties of blue phosphorus. \citet{Xiao2015_01} predict blue phosphorus to be an indirect \textit{p}-type semiconductor with anisotropic properties. They calculate the carrier mobility of a monolayer blue phosphorus under uniaxial and biaxial strains and show strain engineering can tune the properties of blue phosphorus.\\
\citet{Cooper2013_01} use Taylor expansion of the strain energy in order to describe molybdenum disulfide. \citet{Setoodeh2017_01} propose an anisotropic continuum model for black phosphorus. They give two parameter sets for uniaxial stretch along the armchair and zigzag directions. This model does not satisfy periodicity of the black phosphorus lattice and does not predict the behaviour of the structure for loading along an arbitrary direction. So it cannot be used in continuum models such as the membrane model of \citet{Sauer2014_01} and the shell model of \citet{Duong2017_01}. \citet{Kumar2014_01} propose an anisotropic, hyperelastic membrane material model for graphene. This model is nonlinear and directly calibrated from DFT data. It therefore does not require an interatomic interaction potential, and thus avoids any inaccuracies resulting from it. \citet{Kumar2014_01} use their model to simulate nano indentation of a micro scale graphene sheet. \citet{Ghaffari2018_01} extend their model to a Kirchhoff-Love shell model and implement it within the finite element formulation of \citet{Duong2017_01}. Indentation and peeling of graphene sheets, and torsion and bending of carbon nanotubes (CNTs) are simulated with this shell model, and it is used for the nonlinear modal analysis of graphene sheets and CNTs by \citet{Ghaffari2018_02}. \citet{Ghaffari2018_03} propose a new efficient computational shell model for graphene and apply it to carbon nanocones (CNCs). \citet{Shirazian2018_01} propose a new set of material constants for the membrane material model of \citet{Kumar2014_01}, \citet{Ghaffari2018_01} and \citet{Ghaffari2018_03}.\\
In the current work, a new hyperelastic membrane material model is proposed for blue phosphorus. It is based on a set of invariants that are obtained from the symmetry of the lattice. This material model is nonlinear, anisotropic, and calibrated with DFT data. In summary, the novelties of the current work are:
\begin{itemize}
  \item A new hyperelastic continuum membrane material model is proposed for blue phosphorus that is directly based on \textit{ab-initio} data. It thus circumvents the use of atomistic potentials.
  \item It is fully nonlinear and can capture anisotropic behaviour of the material.
  \item Due to the inherent efficiency of continuum formulations, the new model can be used to simulate systems a large length scales.
  \item This is demonstrated by simulating the indentation and vibration of phosphorene at the micrometer scales.
  \item The new model admits extension to other 2D materials, and it can be extended to finite temperatures based on the new formulation of \citet{Ghaffari2019_01}.
\end{itemize}
The remainder of this paper is organised as follows: In Sec.~\ref{s:Surf_Kinematics} the kinematics of deforming surfaces is summarised. In Sec.~\ref{s:structural_tensor_BPh}, a suitable structural tensor and a set of invariants are introduced for blue phosphorus. Based on these, a new hyperelastic membrane material model is proposed in Sec.~\ref{s:BPh_material_model}. In Sec.~\ref{s:BPh_elementary_behav}, the model is calibrated and verified with the DFT results considering various test cases. Sec.~\ref{s:BPh_metric_num_result} presents numerical indentation and modal analysis examples. The paper is concluded in Sec.~\ref{s:BPh_conclusion}.
\section{Kinematics}\label{s:Surf_Kinematics}
In this section, the kinematics of deforming surface is summarised.
\begin{figure}[h]
  \centering
    \includegraphics[height=70mm]{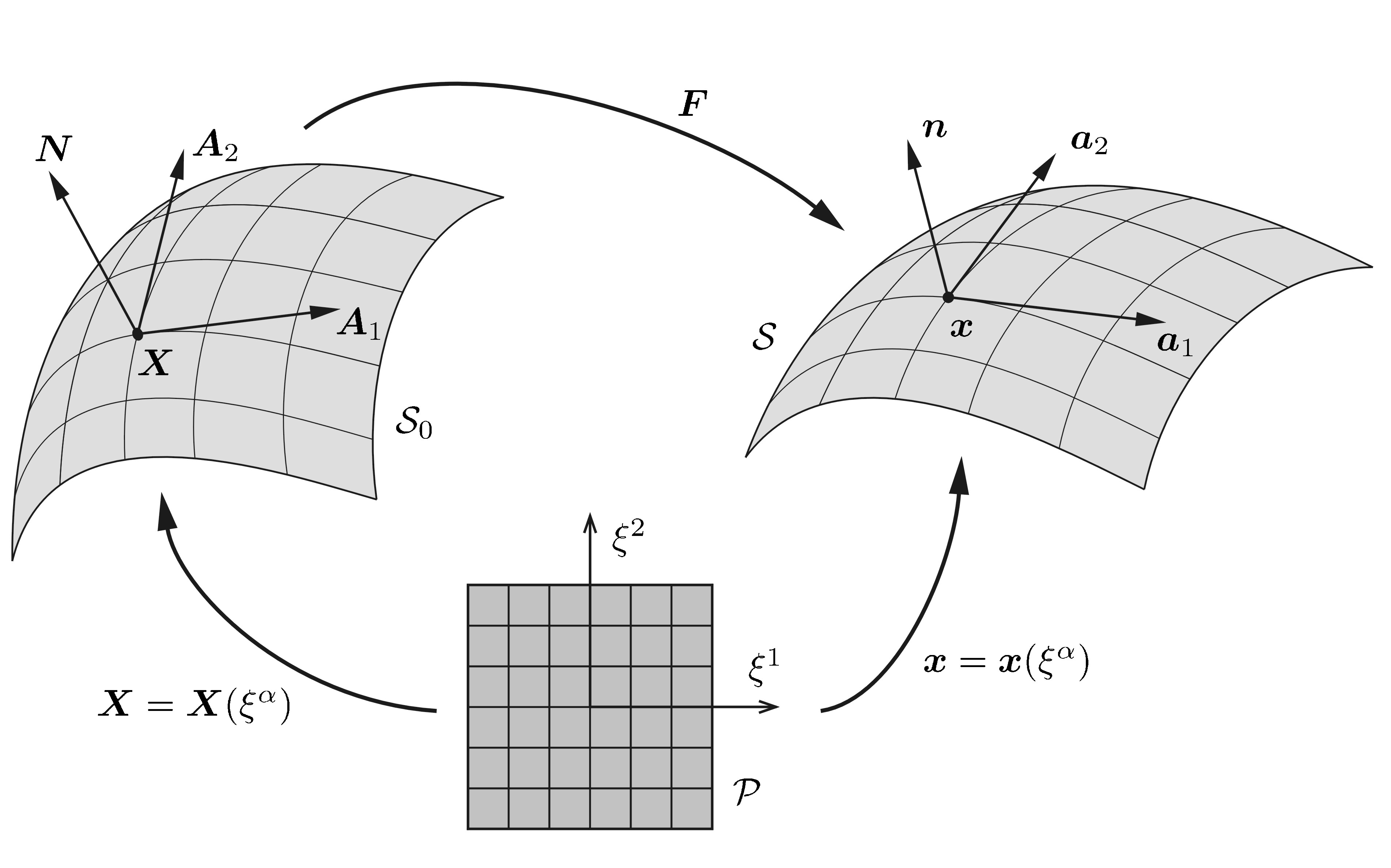}
      \vspace{-3.5mm}
      \caption{Surface description and mapping between $\sS_{0}$ and $\sS$. The figure is adopted from \citet{Sauer2014_01}.\label{f:surface}}
\end{figure}
It will be used in the next sections to propose an anisotropic, hyperelastic continuum material model for blue phosphorus.
\subsection{Surface description}
In order to describe surfaces in very general terms, curvilinear coordinates are used. Accordingly, a point on the surface in the reference configuration $\sS_{0}$ is indicated by
\eqb{lll}
\bX \is \bX(\xi^{\alpha})~,
\eqe
and in the current configuration $\sS$ by
\eqb{lll}
\bx \is \bx(\xi^{\alpha},t)~,
\eqe
see Fig.~\ref{f:surface}. Here $\xi^{\alpha}$, for $\alpha=1,2$, are the curvilinear coordinates and $t$ is time. The tangent vectors of $\sS_{0}$ then follow as
\eqb{lll}
\bA_{\alpha}=\bX_{\!,\alpha}~,
\eqe
and the tangent vectors of $\sS$ as
\eqb{lll}
\ba_{\alpha}=\bx_{,\alpha}~,
\eqe
where ${\bullet,\alpha}=\partial\bullet/\partial\xi^{\alpha}$. The dual vectors $\bA^{\alpha}$ and $\ba^{\alpha}$ are defined such that
\eqb{lll}
\bA_{\alpha}\cdot\bA^{\beta}\is\ba_{\alpha}\cdot\ba^{\beta}=\delta_{\alpha}^{\beta}~,
\eqe
where $\delta_{\alpha}^{\beta}$ is Kronecker delta defined by $[\delta_{\alpha}^{\beta}]=[1~0;0~1]$. The covariant surface metric of $\sS_{0}$ is defined by
\eqb{lll}
\Auab\dis\bA_{\alpha}\cdot\bA_{\beta}~,
\eqe
while the contra-variant surface metric is
\eqb{lll}
\Aab\dis\bA^{\alpha}\cdot\bA^{\beta}~.
\eqe
The covariant, $\auab$, and contra-variant, $\aab$, metric of $\sS$ are defined analogously. The unit normal vector of $\sS_{0}$ is
\eqb{lll}
\bN \is \bA_{1}\times\bA_{2}/\|\bA_{1}\times\bA_{2}\|~,
\eqe
while the unit normal vector of $\sS$ is
\eqb{lll}
\bn\is\ba_{1}\times\ba_{2}/\|\ba_{1}\times\ba_{2}\|~.
\eqe
\subsection{Kinematics of deformation}
The material model will be developed based on the logarithmic surface strain. The logarithmic surface strain can be either defined from the surface stretch tensor or the right Cauchy-Green surface deformation tensor, which both follow from the surface deformation gradient. Differential line elements in the reference and current configuration, denoted $\dif\bX$ and $\dif\bx$, are connected by the rank-two surface deformation gradient $\bF=\ba_{\alpha}\otimes\bA^{\alpha}$ as
\eqb{lll}
\ds \dif \bx \is \ds \bF\,\dif \bX~,
\eqe
where $\bF$ is
\eqb{lll}
\ds \bF \is \ds \ds \pa{\bx}{\bX}~.
\eqe
The polar decomposition of $\bF$ can be written as
\eqb{lll}
\bF \is \bR\,\bU~,
\eqe
where $\bR$ and $\bU$ are the surface rotation and right surface stretch tensor. $\bR$ is proper orthogonal, i.e. $\bR^{-1}=\bR^{\text{T}}$ and $\det \bR=1$, while $\bU$ is symmetric. The spectral decomposition of $\bU$ can be written as
\eqb{lll}
\ds \bU \is \ds \sum_{\alpha=1,2}{\lambda_{\alpha}\,\bY_{\!\!\alpha}\otimes\bY_{\!\!\alpha}}~,
\eqe
where $\lambda_{\alpha}$ and $\bY_{\!\!\alpha}$ are the eigenvalues and eigenvectors of $\bU$. The right Cauchy-Green surface deformation tensor is
\eqb{lll}
\ds \bC \is \ds \bF^{\text{T}}\,\bF=\bU^2=\sum_{\alpha=1,2}{\Lambda_{\alpha}\,\bY_{\!\!\alpha}\otimes\bY_{\!\!\alpha}}~,
\eqe
where $\Lambda_{\alpha}=\lambda^{2}_{\alpha}$, and the eigenvectors of $\bC$ and $\bU$ are the same. $\bU$ or $\bC$ can be used to define the logarithmic surface strain $\bE^{(0)}$ as
\eqb{lll}
\ds \bE^{(0)} \dis \ds \ln\, \bU=\frac{1}{2}\ln\,\bC=\sum_{\alpha=1,2}{\ln(\lambda_{\alpha})\,\bY_{\!\!\alpha}\otimes\bY_{\!\!\alpha}}~.
\eqe
$\bE^{(0)}$ can be additively decomposed into the area-changing part $\bE^{(0)}_{\text{area}}$ and area-preserving part $\bE^{(0)}_{\text{dev}}$ as
\eqb{lll}
\ds \bE^{(0)} \is \ds \bE^{(0)}_{\text{area}}+ \bE^{(0)}_{\text{dev}}~,
\eqe
with
\eqb{lll}
\ds \bE^{(0)}_{\text{area}} \is \ds \frac{1}{2}\tr\left(\bE^{(0)}\right)\bI
\eqe
and
\eqb{lll}
\ds \bE^{(0)}_{\text{dev}} \is \ds \bE^{(0)}-\bE^{(0)}_{\text{area}}~,
\eqe
where $\bI$ is the surface identity tensor on $\sS_{0}$.
\section{Structural tensor and invariants of blue phosphorus}\label{s:structural_tensor_BPh}
\begin{figure}[h]
	\centering
	\includegraphics[height=80mm]{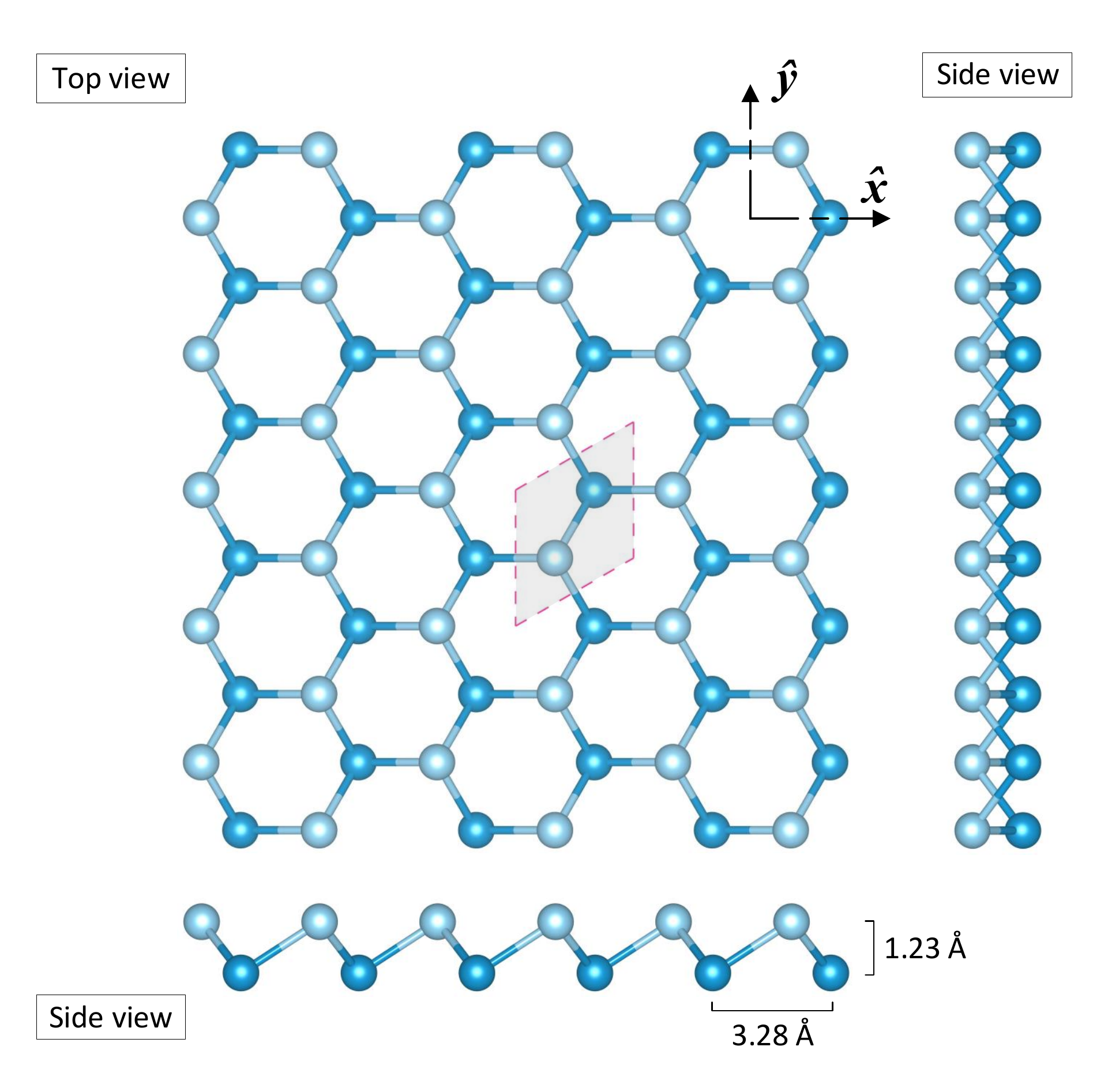}
	\vspace{-3.5mm}
	\caption{Atomic structure of blue phosphorus ($\beta$-P). The primitive cell is highlighted.\label{f:BPh_atomic_Structure}}
\end{figure}

In this section, the structural tensor and a set of invariants for hexagonal structures such as blue phosphorus are given. They are needed in order to model the anisotropic behaviour of the material. The structural tensors of a lattice with a symmetry group of $n$-fold rotational symmetry and reflection plane $C_{nv}$ can be written as \citep{Zheng1993_01}
\eqb{lll}
\ds \bbH_{n}\dis \Re\left[\left(\widehat{\bx}+i\widehat{\by}\right)^{(n)}\right]= \ds \left\{
{\begin{array}{*{20}{ll}}
\Re\left[\left(\widehat{\bM}+i\widehat{\bN}\right)^{(m)}\right]; & n=2m~,\\[3mm]
\Re\left[\left(\widehat{\bx}+i\widehat{\by}\right)\otimes\left(\widehat{\bM}+i\widehat{\bN}\right)^{(m)}\right];&n=2m+1~,
\end{array}} \right.
\eqe
where $i$ is the unit imaginary number, $(\bullet)^{(n)}=(\bullet)\otimes(\bullet)...(\bullet)$ is tensor product of taking ($\bullet$) $n$ times, $\Re$ indicates the real part, $\widehat{\bx}$ and $\widehat{\by}$ are two orthonormal vectors, where at least one of them is in the symmetry plane of the crystal (see Fig.~\ref{f:BPh_atomic_Structure}), and $\widehat{\bM}$ and $\widehat{\bN}$ are defined as
\eqb{lll}
  \widehat{\bM} \dis \widehat{\bx}\,\otimes\,\widehat{\bx}-\widehat{\by}\,\otimes\,\widehat{\by}~,\\[3mm]
  \widehat{\bN} \dis \widehat{\bx}\,\otimes\,\widehat{\by}+\widehat{\by}\,\otimes\,\widehat{\bx}~.
\eqe
For blue phosphorus, $\widehat{\bx}$ is assumed to be in the armchair direction. The blue phosphorus lattice has a rotational symmetry of $2\pi/3$. But, as will be shown later, its constitutive law has the higher symmetry of $\pi/3$, i.e. $n=6$.
In this case, the material invariants based on the logarithmic surface strain can be obtained as \citep{Kumar2014_01,Ghaffari2018_03}
\eqb{lll}
 \ds \sJ_{1} \is \ds \epsilon_{\text{a}}= \ln J~,\\[3mm]
 \ds \sJ_{2} \is \ds \frac{1}{2}\bE^{(0)}_{\text{dev}}:\bE^{(0)}_{\text{dev}}= (\ln\lambda)^2~,\\[3mm]
 \ds \sJ_{3} \is \ds (\ln\lambda)^3\,\cos(6\theta)~,
\eqe
with $\lambda=\sqrt{\lambda_1/\lambda_2}$, where $\lambda_1$ is the larger eigenvalue ($\lambda_1\geqslant\lambda_2$), and $\theta=\arccos(\widehat{\bx}\cdot\bY_{\!\!1})$, where $\bY_{\!\!1}$ is the eigenvector corresponding to the largest eigenvalue. $\theta$ is the maximum stretch angle relative to the armchair direction. $\sJ_{1}$ and $\sJ_{2}$ capture isotropic behaviour and $\sJ_{3}$ captures anisotropic behaviour. $\sJ_{1}$ is characterising area-changing deformations, while $\sJ_{2}$ is characterising shape-changing deformations.
\section{Material model}\label{s:BPh_material_model}
In this section, the previously obtained invariants are used to propose a functional for the strain energy density $W$. From $W$, the Cauchy surface\footnote{with units of [N/m] and in plane-stress format.} stress tensor follows as
\eqb{lll}
\bsig=\ds\frac{1}{\det\bF}\,\bF\,\bS\,\bF^{\text{T}}~,
\label{e:BPh_sig}
\eqe
where
\eqb{lll}
\bS=\ds 2\pa{W}{\bC}~
\eqe
is the second Piola-Kirchhoff surface stress tensor.
In general, $W$ contains several material constants that need to be calibrated from experimental, atomistic or quantum data. In the current work, $W$ is calibrated from DFT results of the strain energy density and Cauchy surface stress tensor $\bsig$. Details on the DFT simulations are given in \ref{s:BPh_DFT_simulations}. A pure dilatation test, and two uniaxial stretching tests (one along the armchair and one along the zigzag direction) are used for the calibration. The material parameters are obtained such that the cost function
\eqb{lll}
\ds \chi \is \ds \sum_{I=1,N_{\text{QM}}} {\|W^{I}-W_{\text{QM}}^{I} \|_{2} + \|\bsig^{I}-\bsig_{\text{QM}}^{I} \|_{2}}~,
\label{e:cost_function}
\eqe
is minimised. Here, ``QM'' indicates the quantum data, $N_{\text{QM}}$ is the number of the data points and $\|\bullet\|_{2}$ is the L2-norm. In this work, the strain energy functional, per unit reference area, is proposed as
\eqb{lll}
\ds W \is \ds f_{1}(\sJ_1)+f_{2}(\sJ_1)\,\sJ_2+f_{3}(\sJ_1)\,\sJ_{2}^{2}+f_{4}(\sJ_1)\,\sJ_{3}~,
\eqe
where $f_{i}$ are polynomial functions of $\sJ_1$ defined as
\eqb{lll}
\ds f_{1} \is \ds n_2\,\sJ_{1}^2+n_3\,\sJ_{1}^3+n_4\,\sJ_{1}^4+n_5\sJ_{1}^5~,\\[3mm]
\ds f_{2} \is \ds \mu_{10}+ \mu_{12}\,\sJ_{1}^2+\mu_{13}\,\sJ_{1}^3+\mu_{14}\,\sJ_{1}^4~,\\[3mm]
\ds f_{3} \is \ds \mu_{20}+\mu_{22}\,\sJ_{1}^2+\mu_{23}\,\sJ_{1}^3+\mu_{24}\,\sJ_{1}^4~,\\[3mm]
\ds f_{4} \is \ds \eta_{0}+\eta_{2}\,\sJ_{1}^2+\eta_{3}\,\sJ_{1}^3+\eta_{4}\,\sJ_{1}^4~.
\eqe
Here $n_i$, $\mu_{ij}$, $\eta_{i}$ are material parameters. The three test cases used for their determination are presented in Sec.~\ref{s:BPh_elementary_behav}. The result of this calibration step are the material parameters given in Tab.~\ref{t:PhBlu_membrane_material_cons_dila}, \ref{t:PhBlu_membrane_material_cons_ish} and \ref{t:PhBlu_membrane_material_cons_ansi}.
\begin{table}[h]
\centering
\caption{Membrane material constants: Pure dilatation part.\label{t:PhBlu_membrane_material_cons_dila}}
\begin{tabular}{lcccc}
  \hline
  & $n_2$ & $n_3$ & $n_4$ & $n_5$ \\ [1mm]
  \hline
  [N/m] & $24.49$ & $-66.07$ & $276.19$ & $-444$ \\[1mm]
  \hline
\end{tabular}
\end{table}
\begin{table}[h]
\centering
\caption{Membrane material constants: Isotropic shear part.\label{t:PhBlu_membrane_material_cons_ish}}
\begin{tabular}{lcccccccc}
  \hline
  & $\mu_{10}$ & $\mu_{12}$& $\mu_{13}$ & $\mu_{14}$ & $\mu_{20}$ & $\mu_{22}$& $\mu_{23}$ & $\mu_{24}$\\ [1mm]
  \hline
  [N/m] &$61.88$ & $-346.44$ & $-670.88$ & $559.77$ & $-1029.84$ & $487.59$ & $-1076.74$ & $-708.98$\\[1mm]
  \hline
\end{tabular}
\end{table}
\begin{table}[h]
\centering
\caption{Membrane material constants: Anisotropic shear part.\label{t:PhBlu_membrane_material_cons_ansi}}
\begin{tabular}{lcccc}
  \hline
  & $\eta_{0}$ & $\eta_{2}$& $\eta_{3}$ & $\eta_{4}$ \\ [1mm]
  \hline
  [N/m] & $62.22$ & $-838.95$ & $-161.41$ & $-427.24$ \\[1mm]
  \hline
\end{tabular}
\end{table}
Using the partial derivatives of the three invariants $\sJ_{i}$ with respect to $\bE^{(0)}$, see \ref{s:derivative_log_inv}, the logarithmic surface stress tensor, $\bS^{(0)}$, that is conjugate to $\bE^{(0)}$, can be computed as
\eqb{lll}
\ds \bS^{(0)} \dis \ds \pa{W}{\bE^{(0)}}= \left[f_{1}^{\prime}+f_{2}^{\prime}\,\sJ_{2}+f_{3}^{\prime}\,\sJ_{2}^{2}+f_{4}^{\prime}\,\sJ_{3}\right]\,\bI+ f_{2}\,\bE^{(0)}_{\text{dev}}+ 2f_{3}\,\sJ_{2}\,\bE^{(0)}_{\text{dev}}+f_{4}\,\pa{\sJ_{3}}{\bE^{(0)}}~,
\label{e:BPh_log_stress}
\eqe
where $f_{i}^{\prime}:=\partial f_{i}/\partial\sJ_{1}$ are
\eqb{lll}
\ds f_{1}^{\prime} \is \ds 2n_2\,\sJ_{1}+3n_3\,\sJ_{1}^2+4n_4\,\sJ_{1}^3+5n_5\,\sJ_{1}^4~,\\[3mm]
\ds f_{2}^{\prime} \is \ds 2\mu_{12}\,\sJ_{1}+3\mu_{13}\,\sJ_{1}^2+4\mu_{14}\,\sJ_{1}^3~,\\[3mm]
\ds f_{3}^{\prime} \is \ds 2\mu_{22}\,\sJ_{1}+3\mu_{23}\,\sJ_{1}^2+4\mu_{24}\,\sJ_{1}^3~,\\[3mm]
\ds f_{4}^{\prime} \is \ds 2\eta_{2}\,\sJ_{1}+3\eta_{3}\,\sJ_{1}^2+4\eta_{4}\,\sJ_{1}^3~.
\eqe
Using the second partial derivatives of $\sJ_{i}$ with respect to $\bE^{(0)}$, see \ref{s:derivative_log_inv}, the corresponding elasticity tensor can be written as
\eqb{lll}
\ds \bbC^{(0)} := \ds \paqq{W}{\bE^{(0)}}{\bE^{(0)}} \is \ds \left[f_{1}^{\prime\prime}+f_{2}^{\prime\prime}\,\sJ_{2}+f_{3}^{\prime\prime}\,\sJ_{2}^{2}+f_{4}^{\prime\prime}\,\sJ_{3}\right]\bI\otimes\bI
+ 2f_{3}\,\bE^{(0)}_{\text{dev}}\otimes\bE^{(0)}_{\text{dev}}\\[3mm]
\plus \ds (f_{2}^{\prime}+2f_{3}^{\prime}\sJ_{2})\,\left[ \bI\otimes\bE^{(0)}_{\text{dev}}+\bE^{(0)}_{\text{dev}}\otimes\bI\right]+f_{4}\,\paqq{\sJ_{3}}{\bE^{(0)}}{\bE^{(0)}}\\[3mm]
\plus \ds f_{4}^{\prime}\,\left[\bI\otimes\pa{\sJ_{3}}{\bE^{(0)}}+\pa{\sJ_{3}}{\bE^{(0)}}\otimes\bI \right]+(f_{2}+2f_{3}\,\sJ_{2})\left[ \sI-\frac{1}{2}\bI\otimes\bI\right]~,
\eqe
where $f_{i}^{\prime\prime}$ are
\eqb{lll}
\ds f_{1}^{\prime\prime} \is \ds 2n_2+6n_3\,\sJ_{1}+12n_4\,\sJ_{1}^2+20n_5\,\sJ_{1}^3~,\\[3mm]
\ds f_{2}^{\prime\prime} \is \ds 2\mu_{12}+6\mu_{13}\,\sJ_{1}+12\mu_{14}\,\sJ_{1}^2~,\\[3mm]
\ds f_{3}^{\prime\prime} \is \ds 2\mu_{22}+6\mu_{23}\,\sJ_{1}+12\mu_{24}\,\sJ_{1}^2~,\\[3mm]
\ds f_{4}^{\prime\prime} \is \ds 2\eta_{2}+6\eta_{3}\,\sJ_{1}+12\eta_{4}\,\sJ_{1}^2~.
\eqe
For a finite element implementation, e.g.~within the model of \citet{Sauer2014_01}, the contra-variant components of the Kirchhoff surface stress tensor $\tauab$ and its corresponding elasticity tensor are needed. Therefore, the logarithmic surface stress and elasticity tensor need be transformed to the Cauchy surface stress tensor $\bsig$. $\tauab$ and the contra-variant components of the Cauchy surface stress tensor, $\sigab$, can be connected as
\eqb{lll}
\tauab\is \det(\bF)\,\sigab~.
\eqe
$\tauab$ can be calculated from the second Piola-Kirchhoff surface stress tensor $\bS$ as
\eqb{lll}
\tauab=\bA^{\alpha}\cdot\bS\bA^{\beta}~,
\eqe
where $\bS$ can be obtained from $W$ using the chain-rule as
\eqb{l}
\bS=\ds 2\pa{W}{\bE^{(0)}}:\pa{\bE^{(0)}}{\bC}~,
\label{e:BPh_PK2}
\eqe
where $\partial W/\partial\bE^{(0)}$ is given in \eqref{e:BPh_log_stress} and $\partial\bE^{(0)}/\partial\bC$ can be found in \citet{Kumar2014_01} and \citet{Ghaffari2018_01}.
The transformation of \eqref{e:BPh_PK2} and the corresponding transformation for the elasticity tensor are provided by \citet{Kumar2014_01} and \citet{Ghaffari2018_01} for graphene. \citet{Ghaffari2018_03} pointed out that this transformation is computationally expensive and rather complicated. Therefore a finite difference formulation is considered here as discussed in \ref{s:Finite_difference_stress_elasticity}.\\
Next, the thickness variation $\lambda_3$ is discussed. $\lambda_3$ is computed from the DFT simulations which are used to propose the following isotropic function based on the first and second invariants
\eqb{lll}
\lambda_{3} \is 1+p_1\,\sJ_{1}+p_2\,\sJ_{1}^7+ s_{2}\,\sJ_{2}^2+ s_{3}\,\sJ_{2}^3  + s_{4}\,\sJ_{2}^4~,
\eqe
where $p_i$ and $s_{i}$ are given in Tab.~\ref{t:PhBlu_thickness_cons}. A cost function similar to \eqref{e:cost_function} is used to determine these constants from the three test cases presented in the following section.
\begin{table}[h]
\centering
\caption{Constants for the thickness variation.\label{t:PhBlu_thickness_cons}}
\begin{tabular}{ccccc}
  \hline
  $p_{1}$ & $p_{2}$& $s_{2}$ & $s_{3}$  & $s_{3}$\\ [1mm]
  \hline
   $-0.22$ & $-27.48$ & $-1245$ & $119$ & $-3.77$ \\[1mm]
  \hline
\end{tabular}
\end{table}
\section{Model calibration and validation}\label{s:BPh_elementary_behav}
In this section, first the three calibration tests are described.. Then, the continuum results from the proposed blue phosphorus material model are compared and verified with DFT results of the current work and the literature. Finally the linear elastic constants are reported and compared with the literature.
\subsection{Calibration}
\begin{figure}[h]
  \centering
      \begin{subfigure}{.49\textwidth}
        \centering
    \includegraphics[height=59mm]{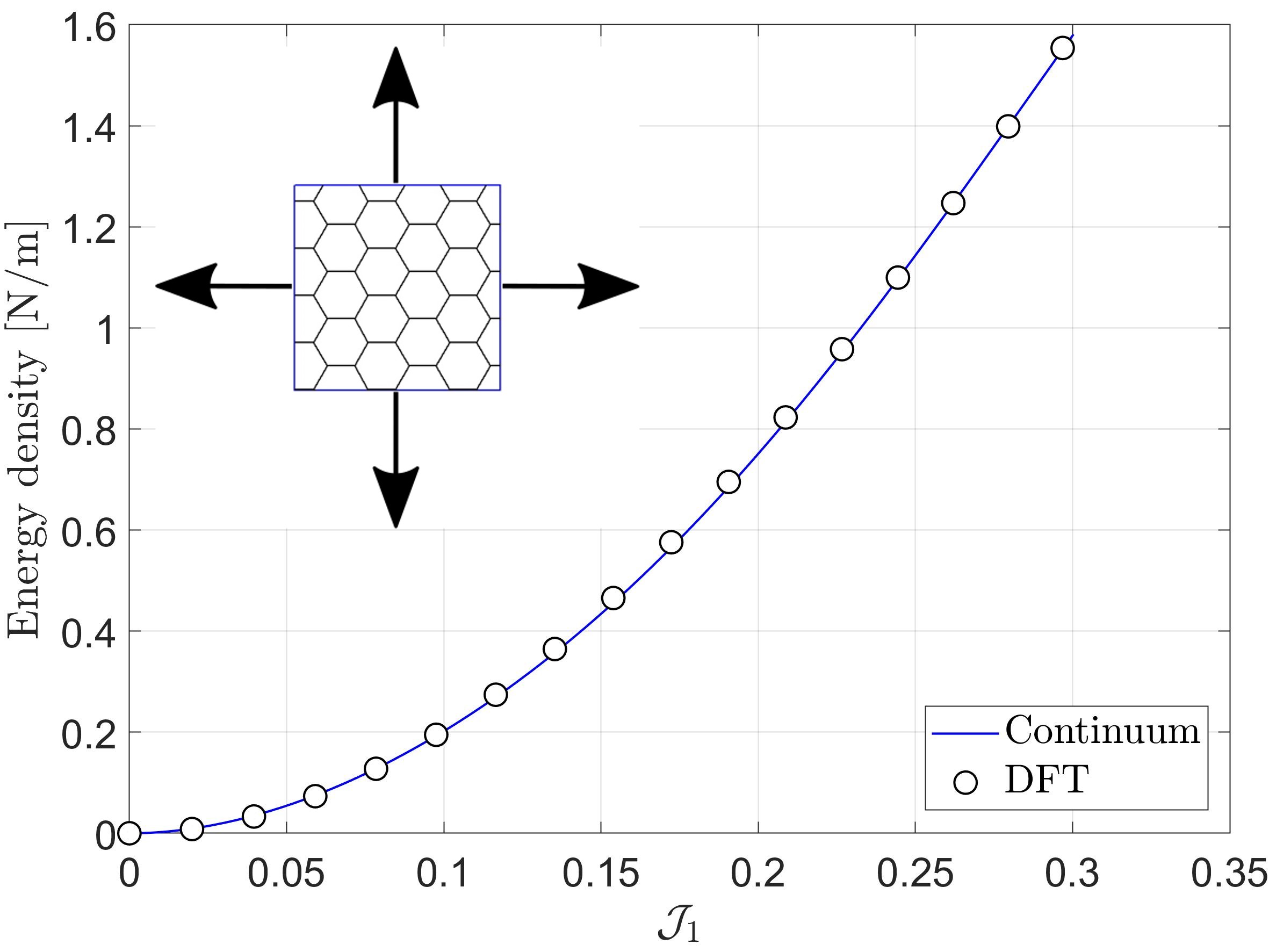}
        \vspace{-7.5mm}
        \subcaption{}
    \end{subfigure}
    \begin{subfigure}{.49\textwidth}
        \centering
    \includegraphics[height=59mm]{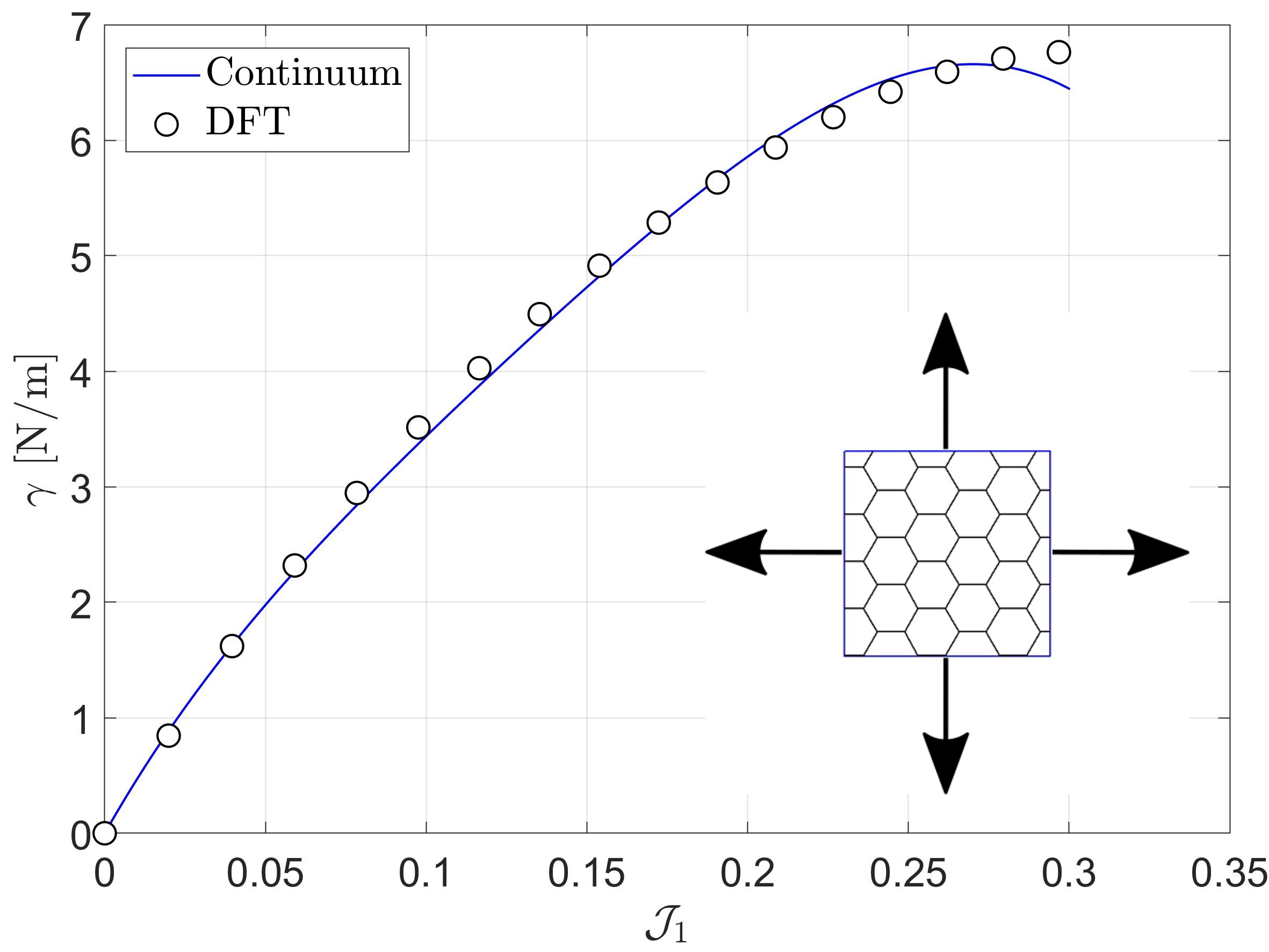}
        \vspace{-7.5mm}
        \subcaption{}
    \end{subfigure}
      \vspace{-3.5mm}
      \caption{Pure dilatation calibration: (a) Strain energy density per unit reference area; (b) surface tension. 163 DFT data sets are used for calibration.\label{f:PhBlu_dilatation_Fr_Comparison}}
\end{figure}

\begin{figure}[h]
  \centering
      \begin{subfigure}{.49\textwidth}
        \centering
    \includegraphics[height=59mm]{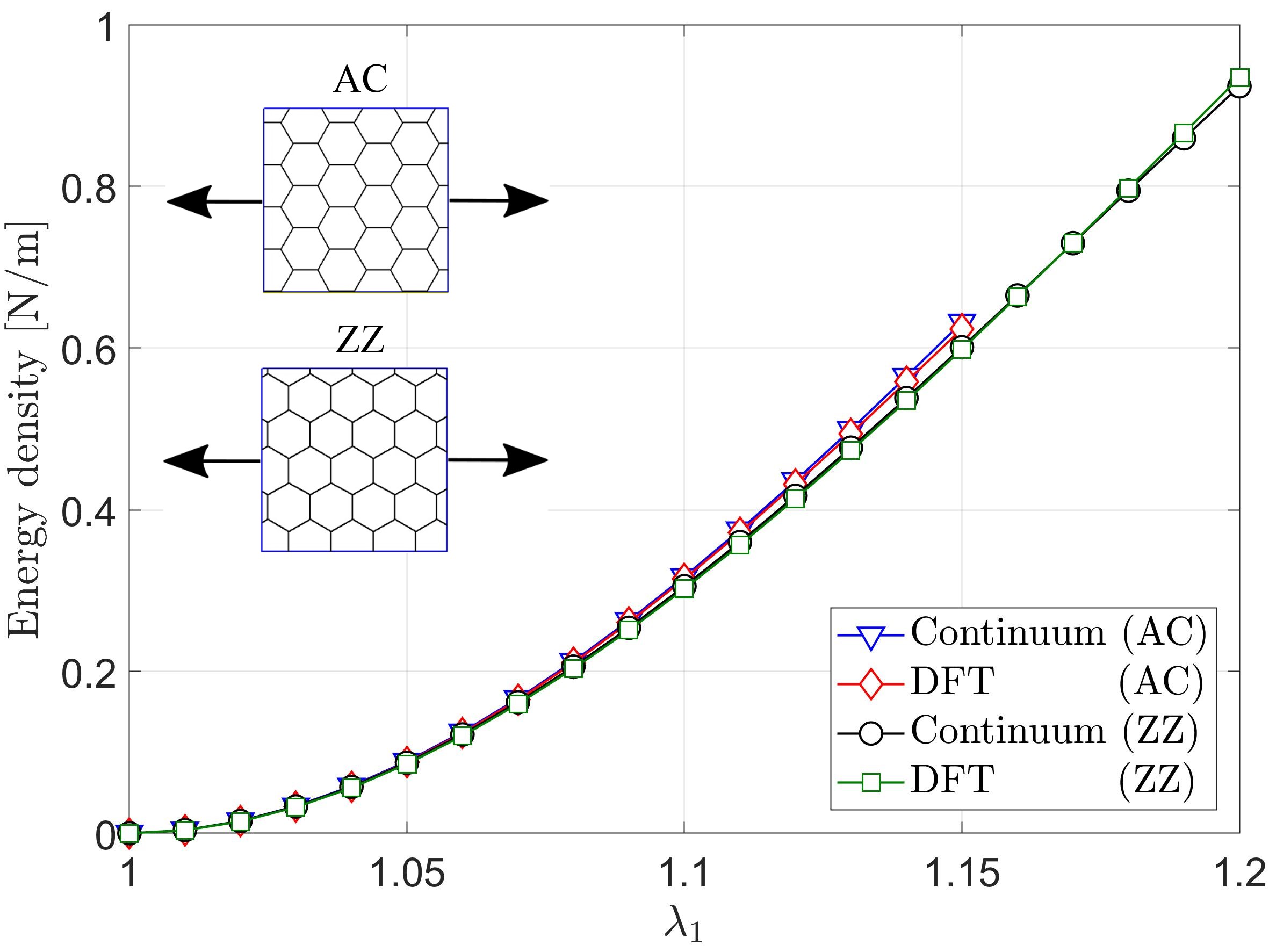}
        \vspace{-7.5mm}
        \subcaption{}
    \end{subfigure}\\
    \begin{subfigure}{.49\textwidth}
        \centering
    \includegraphics[height=59mm]{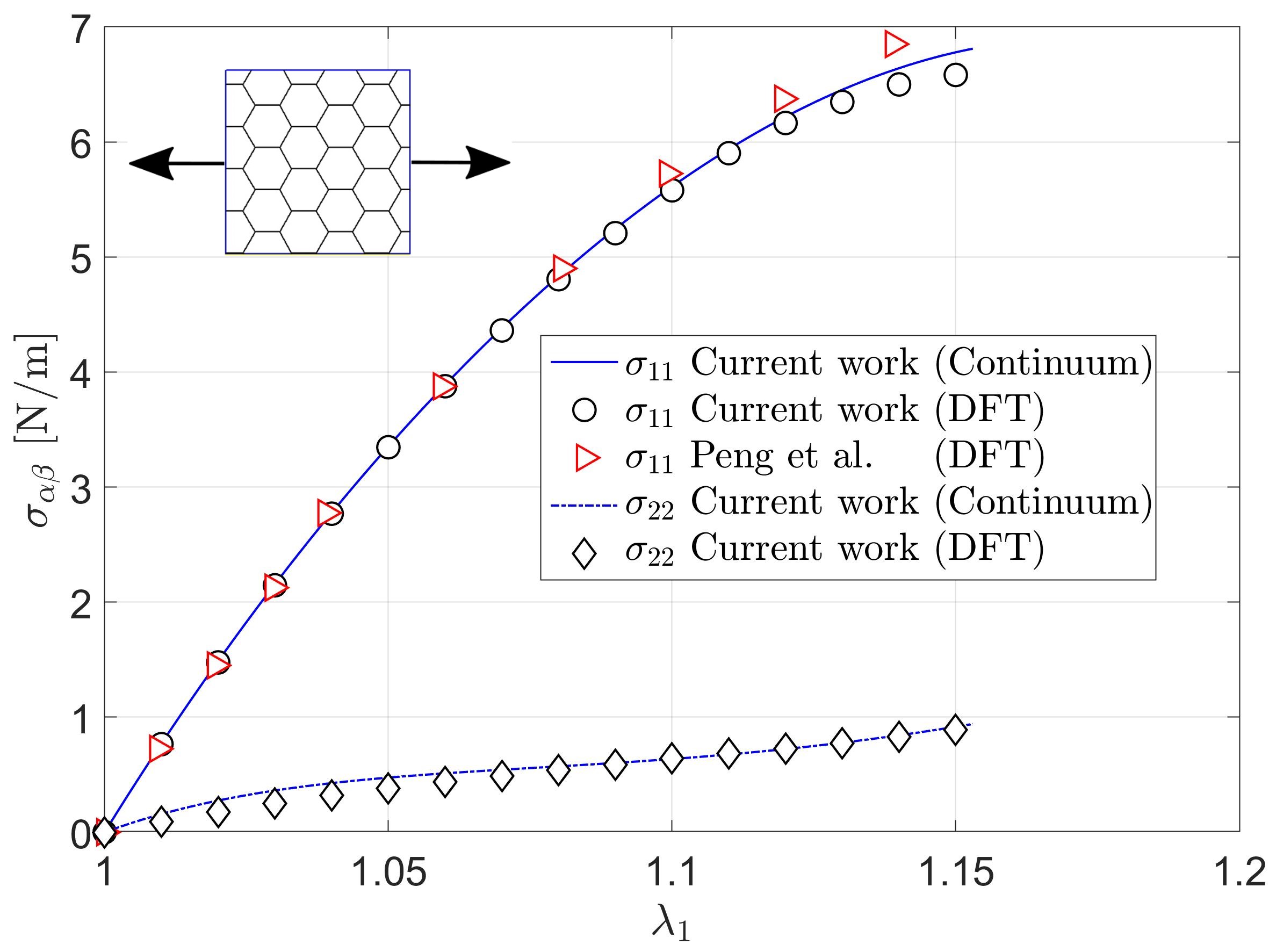}
        \vspace{-7.5mm}
        \subcaption{}
    \end{subfigure}
        \begin{subfigure}{.49\textwidth}
        \centering
    \includegraphics[height=59mm]{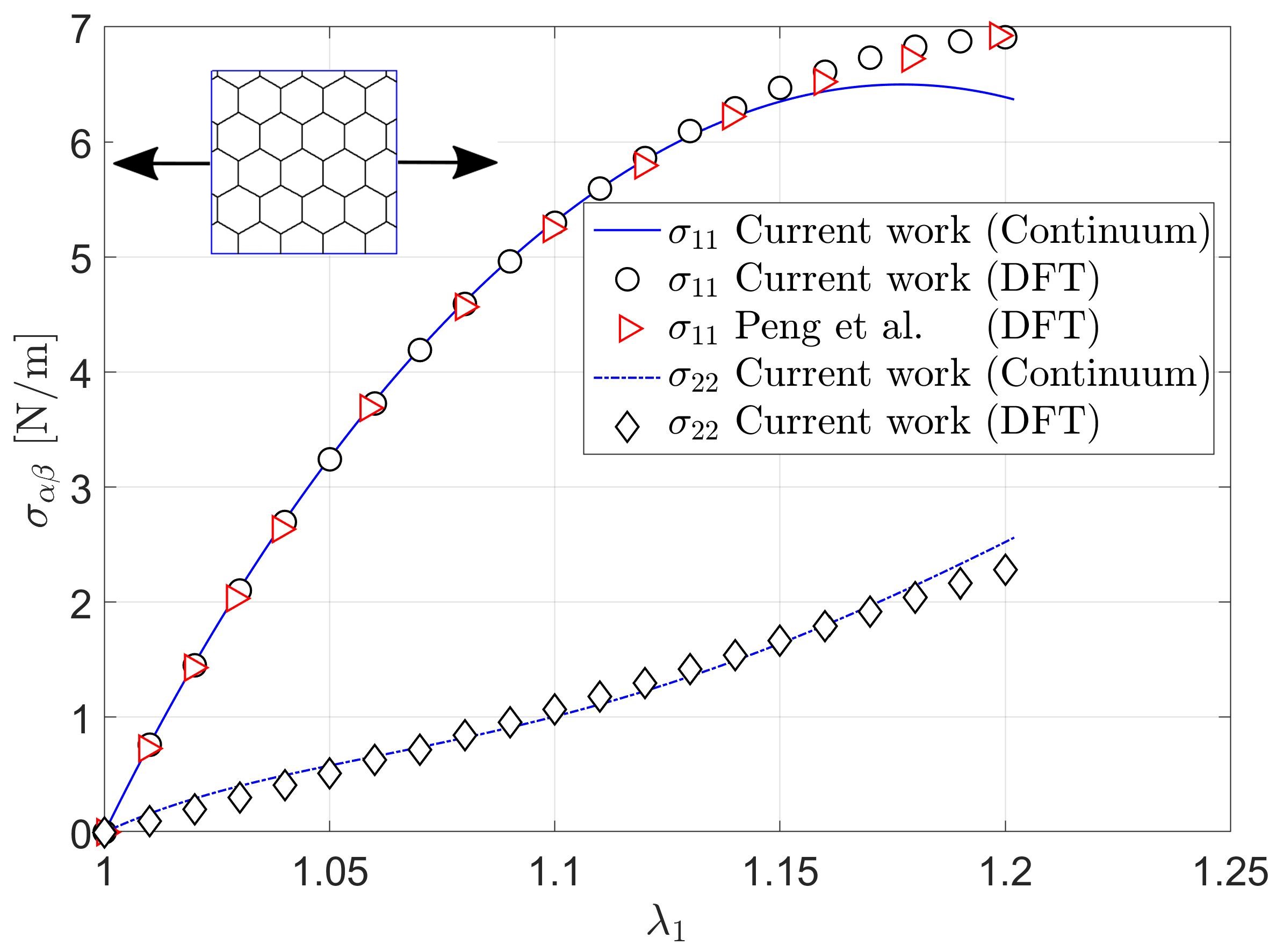}
        \vspace{-7.5mm}
        \subcaption{}
    \end{subfigure}
      \vspace{-3.5mm}
      \caption{Uniaxial stretch calibration and validation for armchair (AC) and zigzag (ZZ) stretch: (a) Strain energy density per unit reference area; Cauchy surface stress for uniaxial stretch along the (b) armchair and (c) zigzag direction. The lattice is deformed in the armchair or zigzag direction and fixed in the perpendicular direction. $\sigma_{11}$ and $\sigma_{22}$ are computed from \eqref{e:BPh_sig} and are the Cartesian components of the Cauchy surface stress in the stretched direction and the direction perpendicular to it, respectively. The current results are also compared with the DFT results from \citet{Peng2016_01}. The calibration is based on 154 and 203 DFT data sets for uniaxial stretch along the armchair and zigzag directions, respectively.\label{f:PhBlu_Fr_uniaxial_comparsion}}
\end{figure}

\begin{figure}[h]
  \centering
      \begin{subfigure}{.49\textwidth}
        \centering
    \includegraphics[height=59mm]{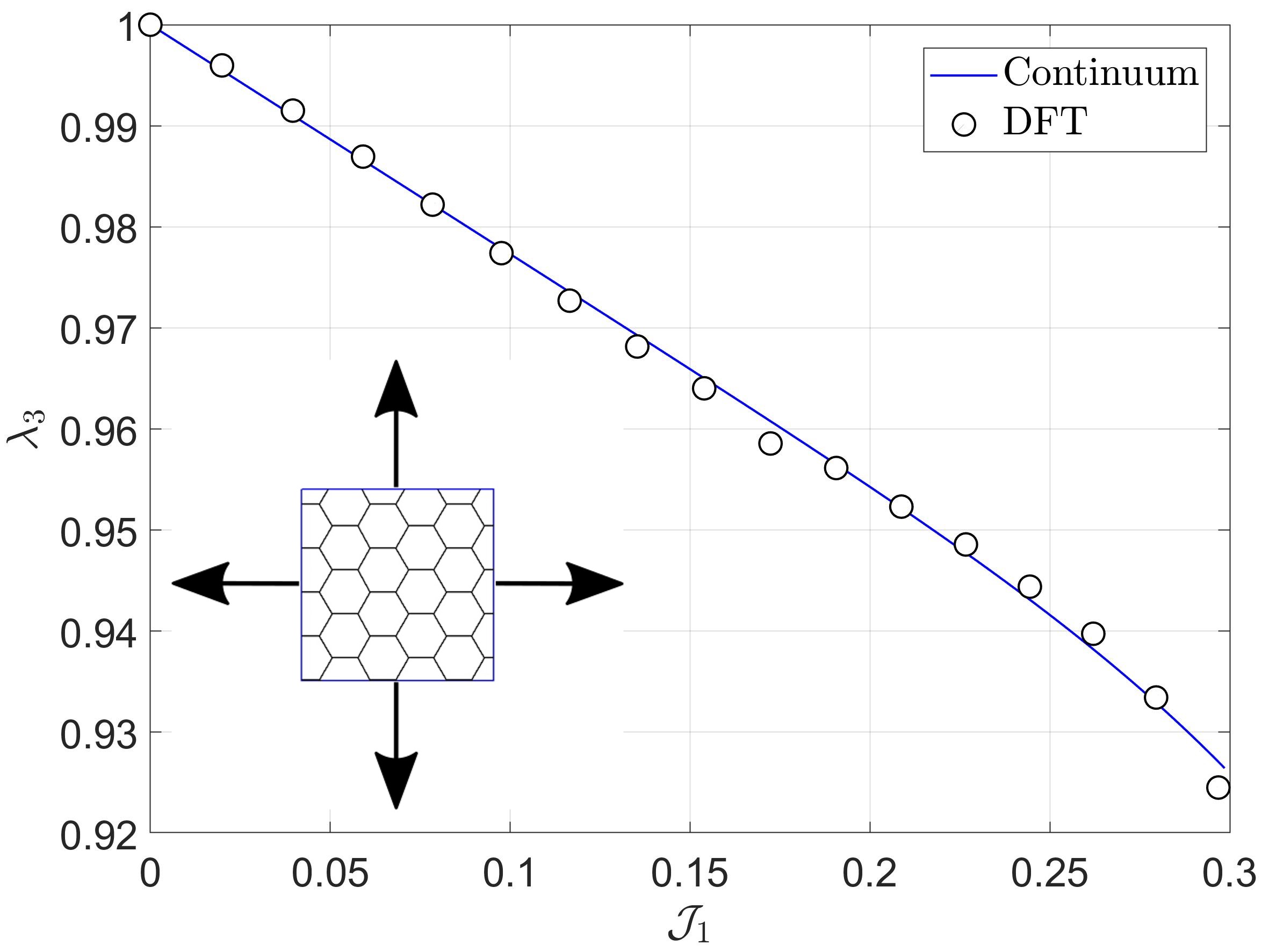}
        \vspace{-7.5mm}
        \subcaption{}
    \end{subfigure}
    \begin{subfigure}{.49\textwidth}
        \centering
    \includegraphics[height=59mm]{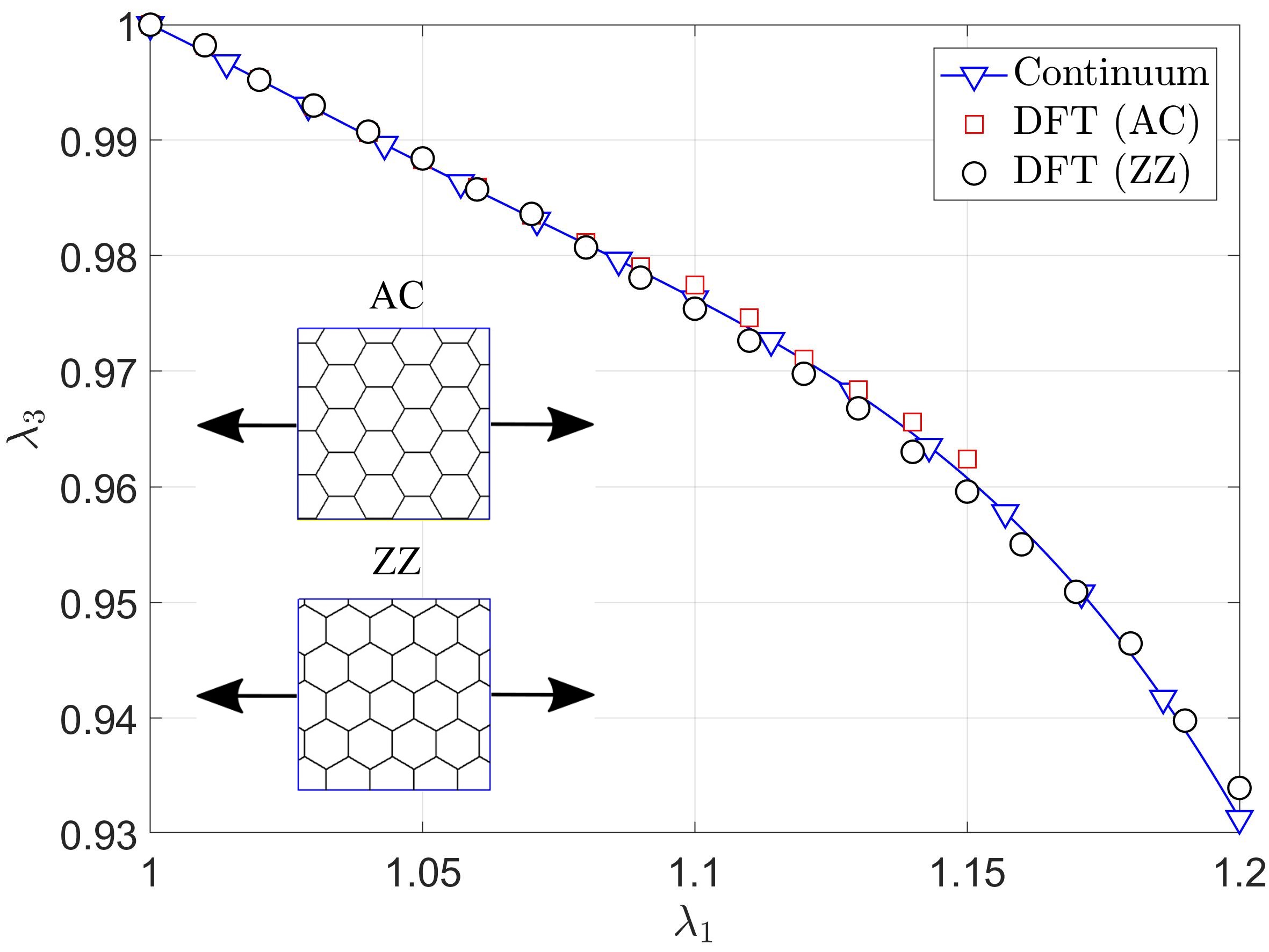}
        \vspace{-7.5mm}
        \subcaption{}
    \end{subfigure}
      \vspace{-3.5mm}
      \caption{Calibration of the thickness change $\lambda_3$: For (a) pure dilatation; (b) uniaxial stretch.  AC = armchair and ZZ = zigzag. The calibration is based on 163, 154 and 203 DFT data sets for pure dilatation and uniaxial stretch along the armchair and zigzag directions, respectively.\label{f:BPh_thickness_variation}}
\end{figure}

The calibration is conducted by performing one pure dilatation test and two uniaxial stretching tests. Altogether, 520 DFT data sets are used for this calibration, leading to the material parameters listed in Tab.~\ref{t:PhBlu_membrane_material_cons_dila}, \ref{t:PhBlu_membrane_material_cons_ish} and \ref{t:PhBlu_membrane_material_cons_ansi}.\\
For pure dilatation, the hexagonal structure does not change and the material behaviour is isotropic. Fig.~\ref{f:PhBlu_dilatation_Fr_Comparison} shows the variation of the energy density and surface tension\footnote{$\gamma=1/2\,\tr\bsig$.} with respect to the area change invariant $\sJ_{1}$.
The energy density increases monotonically. The surface tension reaches a maximum and then begins to decrease. The DFT results are only considered up to this point since the lattice collapses beyond that point. This maximum occurs sooner in the continuum model than in the DFT results.\\
Next the lattice is stretched in the armchair or zigzag direction, while being fixed in the perpendicular direction in order to produce uniaxial stretch. The energy density and Cauchy surface stress components are shown in Fig.~\ref{f:PhBlu_Fr_uniaxial_comparsion}.
The material behaves anisotropic under uniaxial stretch and it fails sooner if stretched in the armchair direction rather than the zigzag direction. The strain energy density and stress $\sigma_{11}$ for stretch along the armchair direction are higher than for stretch along the zigzag direction, while it is vice versa for $\sigma_{22}$. The continuum and DFT results are in good agreement for all tests. The results are reported up to the maximum stress point beyond which the material becomes unstable.\\
Fig.~\ref{f:BPh_thickness_variation} shows the calibration of the out-of-plane stretch $\lambda_{3}$ under pure dilatation and uniaxial stretch. The same 520 DFT data sets are used for this calibration, leading to the parameters listed in Tab.~\ref{t:PhBlu_thickness_cons}.
The material shows an approximate isotropic response for both cases. In both cases $\lambda_{3}$ decreases monotonically up to the failure point.
\subsection{Validation}\label{ss:BPh_validation}
\begin{figure}[h]
	\centering
	\includegraphics[height=59mm]{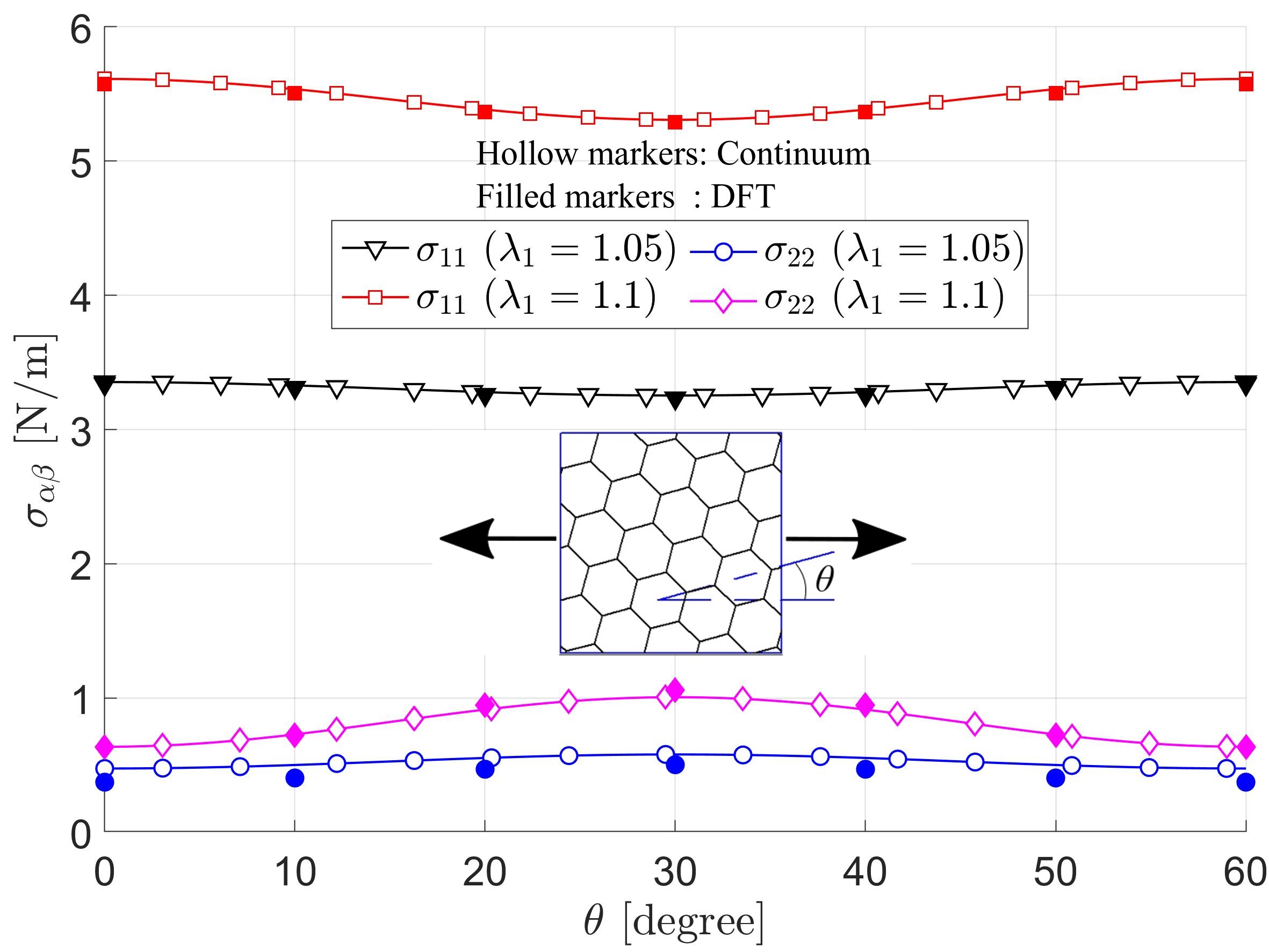}
	\vspace{-3.5mm}
	\caption{Uniaxial stretch validation: Comparison of continuum and DFT results of $\sigma_{11}$ and $\sigma_{22}$ for a stretch in arbitrary direction $\theta$. This is measured with respect to the armchair direction. $\sigma_{11}$ and $\sigma_{22}$ are the Cartesian components of the Cauchy surface stress in the stretch direction and the direction perpendicular to it, respectively.\label{f:PhBlu_Fr_uniaxial_Stress_cart}}
\end{figure}

For uniaxial stretch along the armchair and zigzag direction, the current continuum model and DFT results are compared and validated with the DFT results of \citet{Peng2016_01} in Fig.~\ref{f:PhBlu_Fr_uniaxial_comparsion}. As seen they are a good agreement.\\
For further validation of the model first, the Young's modulus $E$, shear modulus $G$, bulk modulus $K$ and Poisson's ratio $\nu$ are obtained in the small deformation regime and compared with reported values from the literature in Tab.~\ref{t:PhBlu_linear_lam_para}. Given the plane-stress material model in Sec.~\ref{s:BPh_material_model}, those material constant turn out to be $E=4G\,K/(G+K)$, $G=\mu_{10}/2$, $K = 2n_2$ and $\nu=(K-G)/(K+G)$. Compared to other results, the proposed continuum model behaves stiffer by about 10-15\%.\\
\begin{table}[h]
\centering
\caption{Linear elasticity constants: AC = armchair and ZZ = zigzag.\label{t:PhBlu_linear_lam_para}}
\begin{tabular}{lcccc}
  \hline
  & $E$~[N/m] & $G$~[N/m]& $K$~[N/m]& $\nu$ \\ [1mm]
  \hline
  Current work & $75.85$ (in AC and ZZ) & $30.94$ & $48.98$ & $0.226$ \\[1mm]
  \citet{Sun2016_01} & - & - & 42.66 & - \\[2mm]
  \citet{Peng2016_01} & 69.36 (in AC) and 66.24 (in ZZ) & - & - & - \\
  \hline
\end{tabular}
\end{table}\noindent
Next, the material is stretched in different directions and the variation of the stresses with respect to the stretch direction is reported in Fig.~\ref{f:PhBlu_Fr_uniaxial_Stress_cart}. The continuum and DFT results are in very good agreement. The DFT results for $\theta=10,20,40,50^\circ$ have not been used for the material calibration and so they can be used for the validation of the model. The material has a rotational periodicity of $\pi/3$ since a rotation of $\pi/3$ results in a mirror transformation of the structure with respect to its plane, which does not affect the mechanical properties of the blue phosphorus monolayer.
\section{Simulation results at large length scales}\label{s:BPh_metric_num_result}
The model has been validated in the previous section. In this section, it will be used to model a large micro-meter phosphorene specimen under indentation and vibration. The specimen consists of more than 5 million atoms. Molecular dynamics simulations would be too expensive for this length scale. The continuum model, on the other hand, has only 9000 nodes for the finest mesh. Thus the continuum model has more than 500 times less degrees of freedom than the atomistic system. Beyond that, there presently exists no suitable atomistic potential for blue phosphorus.
\subsection{Phosphorene indentation}
\begin{figure}[h]
  \centering
    \begin{subfigure}{0.49\textwidth}
    \includegraphics[width=50mm, angle =90 ]{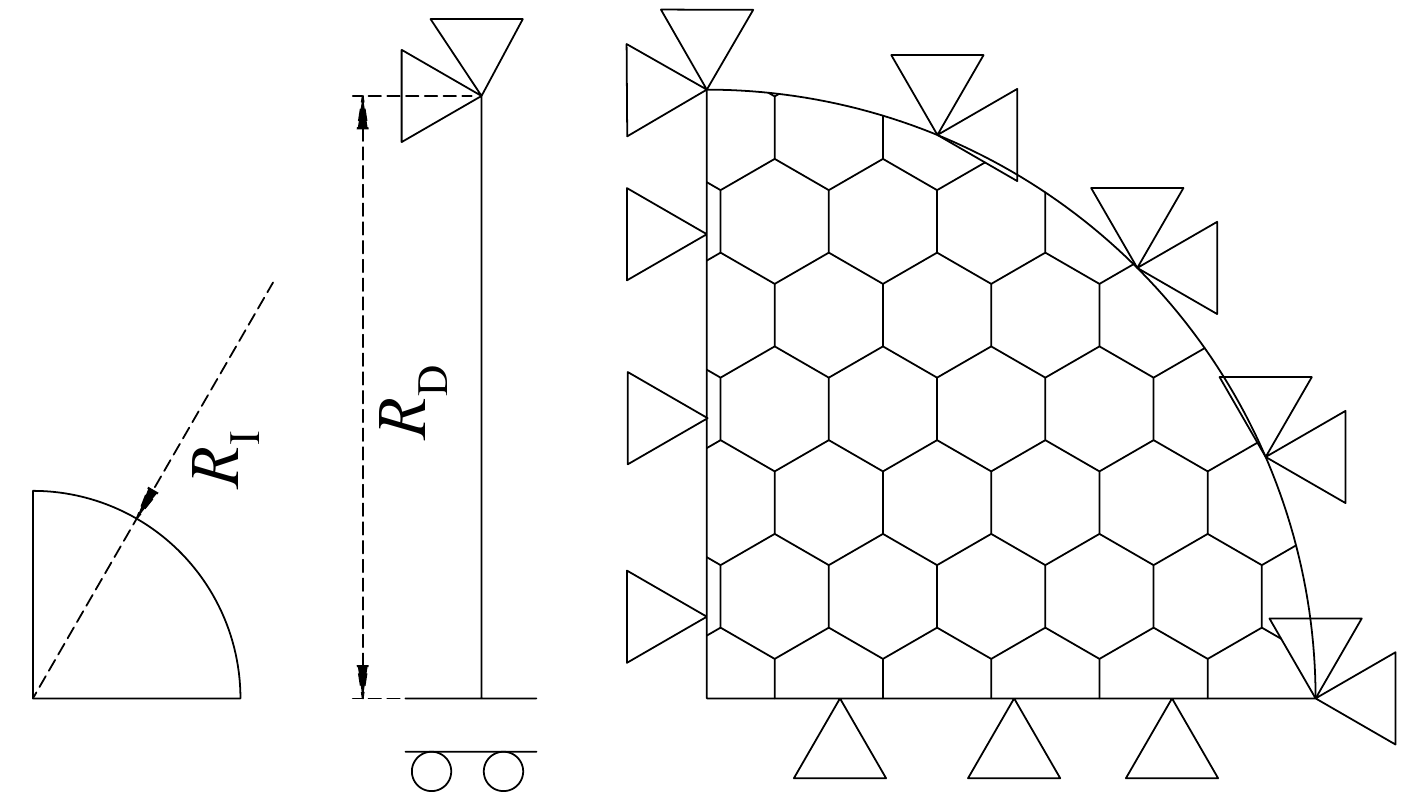}
      \vspace{-10.5mm}
        \subcaption*{(a)}
    \end{subfigure}
         \begin{subfigure}{0.49\textwidth}
        \centering
    \includegraphics[height=50mm,trim=1cm 3cm 6cm 4,clip]{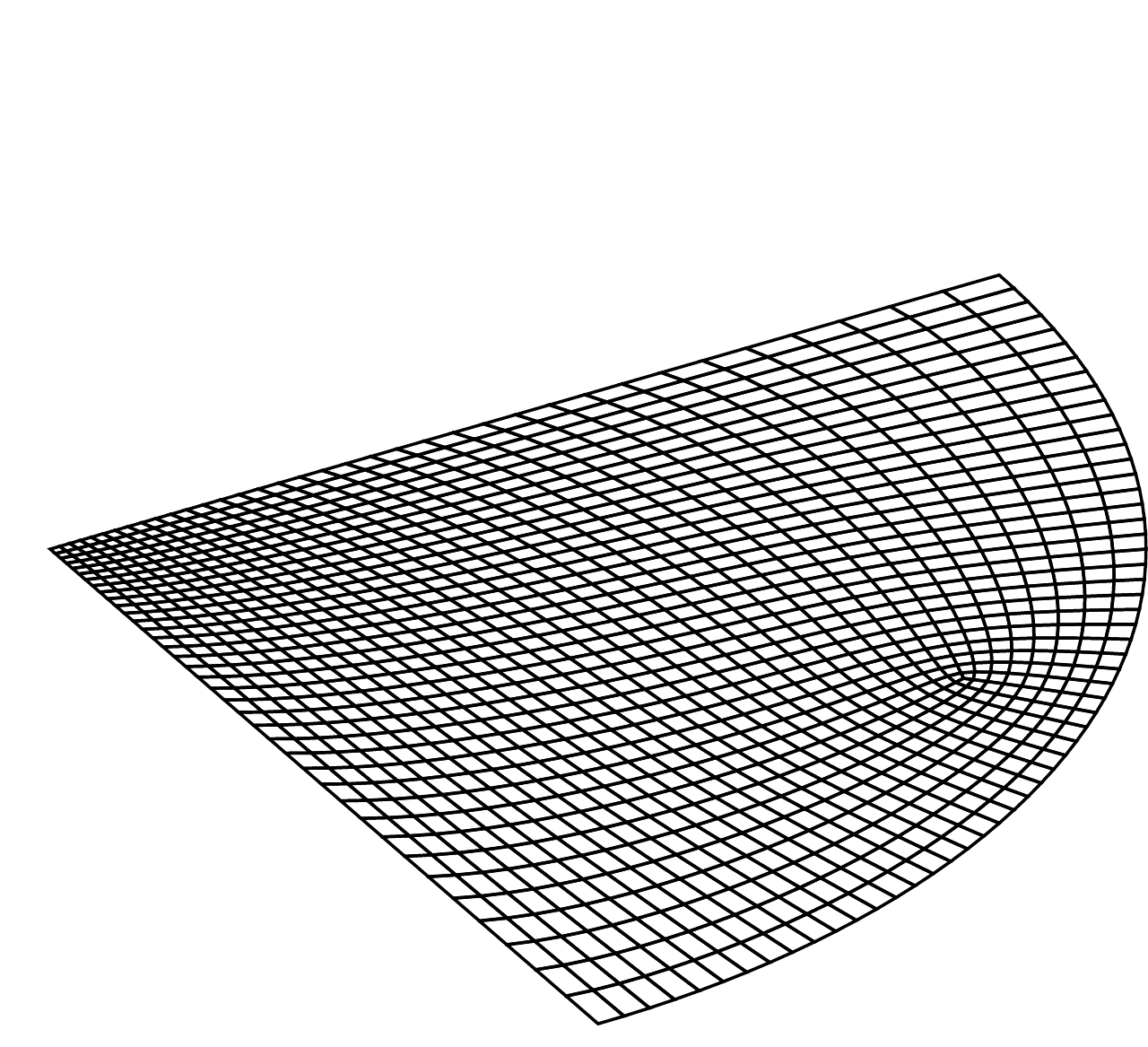}
        \vspace{-3.5mm}
        \subcaption*{(b)}
    \end{subfigure}
    \vspace{-3.5mm}
      \caption{Phosphorene indentation: (a) Boundary conditions; (b) the coarsest finite element mesh. Sheet and indentor radii, $R_{\text{D}}$ and $R_{\text{I}}$, are 500 nm and 16.5 nm, respectively.\label{f:PhBlue_indentation_bc_mesh}}
\end{figure}

\begin{figure}[h]
  \centering
    \includegraphics[width=160mm]{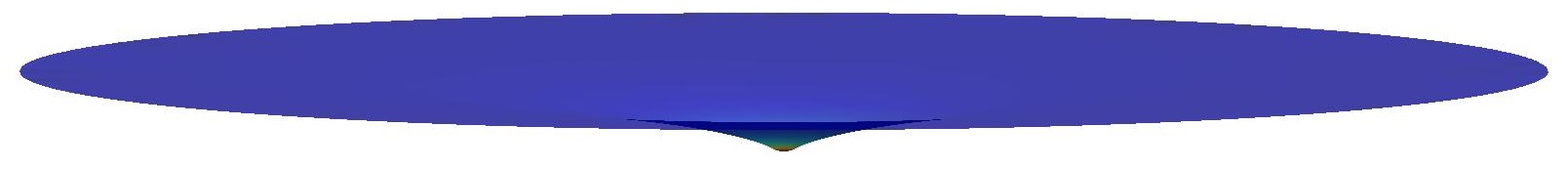}
     \vspace{-5mm}
    \caption{Phosphorene indentation: Deformation for indentor displacement 53 nm and phosphorene pre-strain~1\%.\label{f:BPh_indentation_deformed_geo}}
\end{figure}

\begin{figure}[h]
	\centering
	\includegraphics[height=59mm]{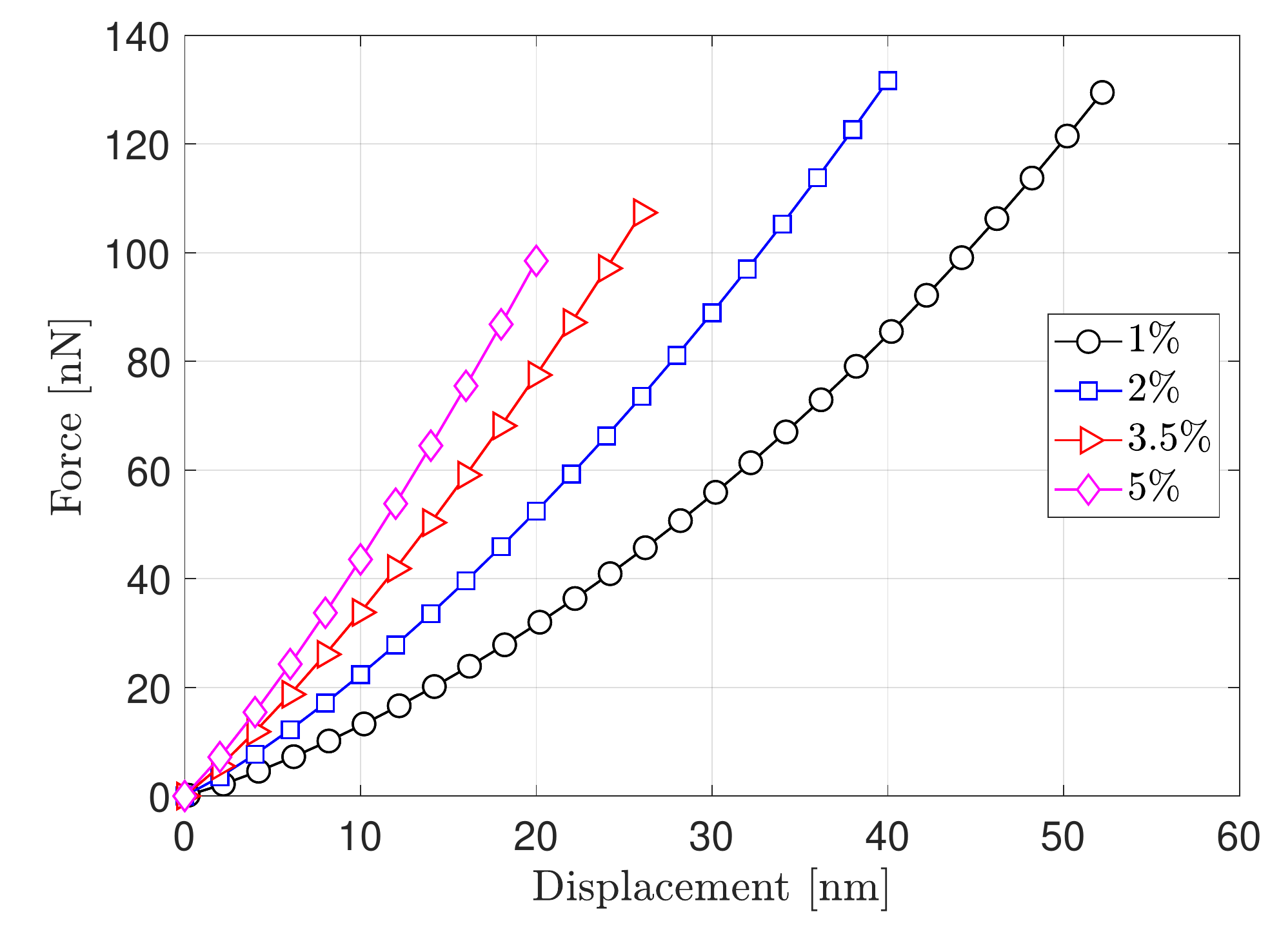}
	\vspace{-5.5mm}
	\caption{Phosphorene indentation: Force-displacement curves for the in-plane pre-strains 1\%, 2\%, 3.5\% and 5\%. \label{f:BPh_Indentation_Force_displacement}}
\end{figure}

A circular blue phosphorus sheet is indented with a spherical indentor. The boundary is fixed (but free to rotate) as shown in Fig.~\ref{f:PhBlue_indentation_bc_mesh}a.
The membrane is pre-strained by different magnitudes in order to stabilize it. For simplicity, only one quarter of the specimen is modeled using the FE formulation of \citet{Sauer2014_01}. The phosphorene specimen and indentor radii are 500 nm and 16.5 nm, respectively. Quadratic Lagrangian meshes with 2536, 2670, 3366, 4462 and 6506 finite elements are used for the convergence study (see Fig.~\ref{f:PhBlue_indentation_bc_mesh}b for the coarsest mesh). The reaction force only changes by about 0.05\% between the finest and second finest mesh. The deformed geometry and force-displacement graph are shown in Figs.~\ref{f:BPh_indentation_deformed_geo} and \ref{f:BPh_Indentation_Force_displacement}.
\noindent
\noindent
The force-displacement curve is linear for high pre-strains but shows pronounced nonlinearity otherwise. The curves terminate at a certain load level due to material instability (i.e. failure).
\subsection{Phophorene vibration}
\begin{figure}[h]
	\centering
	\begin{subfigure}{.49\textwidth}
		\includegraphics[height=59mm]{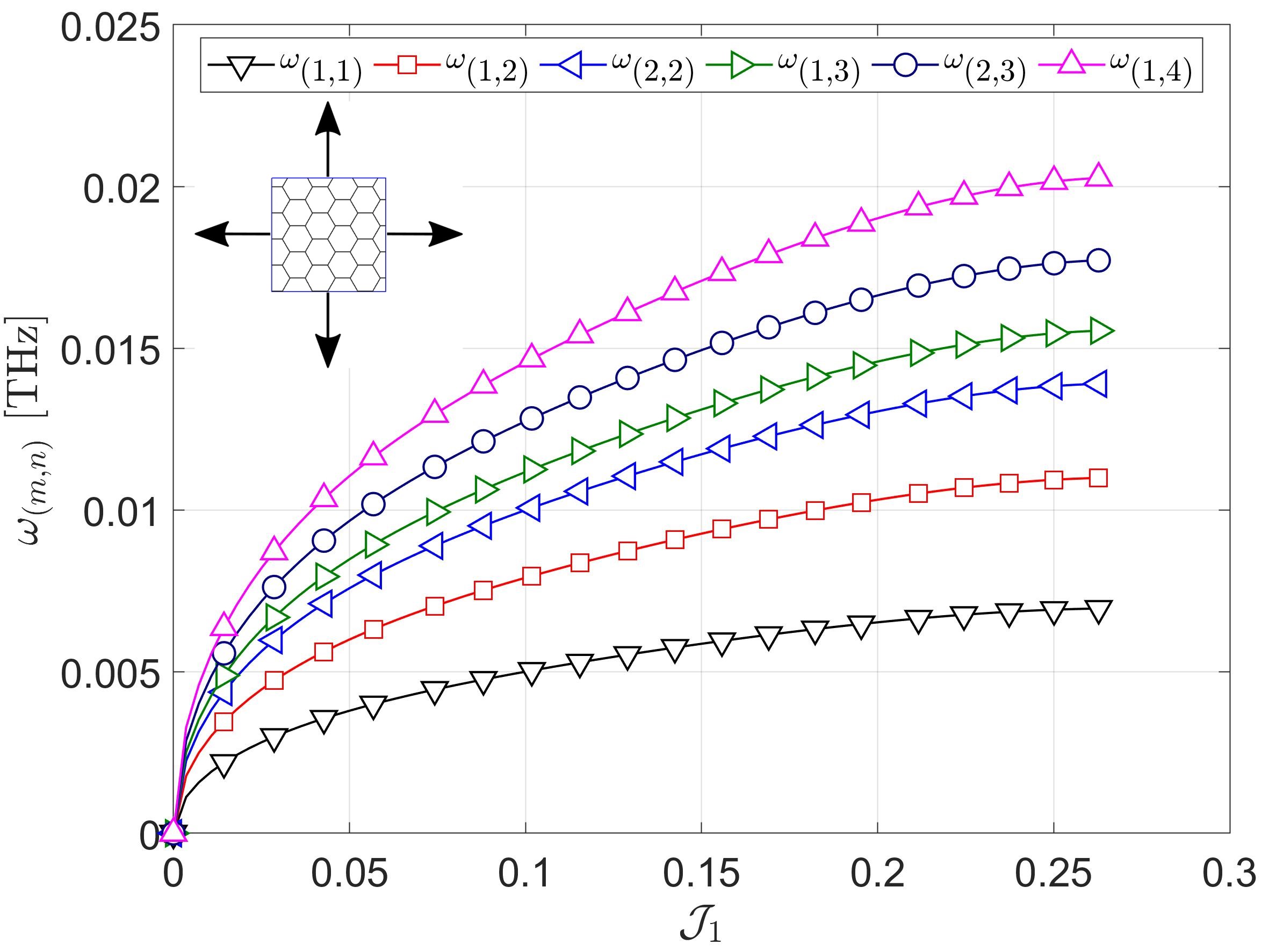}
		\vspace{-7.5mm}
		\subcaption*{(a)}
	\end{subfigure}
	\begin{subfigure}{.49\textwidth}
		\includegraphics[height=59mm]{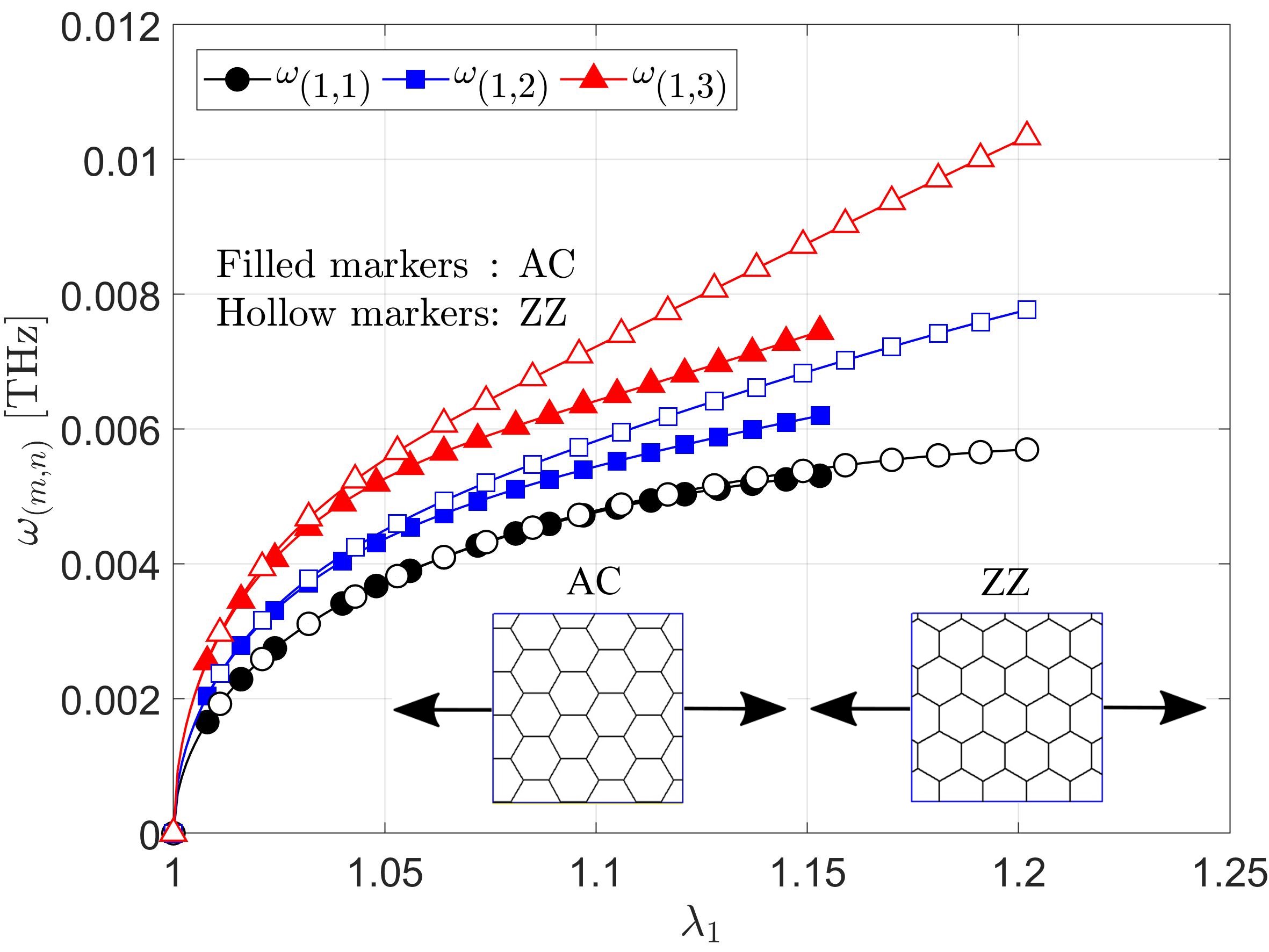}
		\vspace{-7.5mm}
		\subcaption*{(b)}
	\end{subfigure}
	\vspace{-3.5mm}
	\caption{Phosphorene modal analysis: Frequency variation under (a) pure dilatation and (b) uniaxial stretch along the armchair (AC) and zigzag (ZZ) directions. The square membrane sheet has an half edge length of a = b = 250 nm.\label{f:BPh_Rect_simply_freqeuncy_variation}}
\end{figure}

Next, the variation of the frequencies for a square phosphorene sheet is obtained analytically. Since the model is validated in Sec.~\ref{ss:BPh_validation}, we expect the results presented here to be correct. The frequencies of a pre-strained rectangular sheet with edge lengths $a$ and $b$ are \citep{Leissa1969_01,Ghaffari2018_03}
\eqb{lll}
\ds \omega_{(m,n)}^2 \is \ds \hat{\omega}^{2}_{(m,n)}+\frac{1}{\rho}\left[\sigma_{x}\left(\frac{\pi\,m}{ a}\right)^2+\sigma_{y}\left(\frac{\pi\,n}{b}\right)^2\right]~,
\label{e:analytical_sol_rect_nonlinear_modal}
\eqe
where $\omega_{(m,n)}^2$ are the square of the frequencies and $\hat{\omega}^{2}_{(m,n)}$ are related to the bending stiffness, which can be neglected for large sheets ($\hat{\omega}_{(m,n)}\approx 0$) \citep{Ghaffari2018_03}, $\rho=1.101\times10^{-6}$ kg/m$^2$ is the surface mass density, $a$ and $b$ are the half length and half width of the sheet and $\sigma_{x}$ and $\sigma_{y}$ are the normal surface stress components along the edges $a$ and $b$ \citep{Leissa1969_01}. $m$ and $n$ are the number of half waves in the mode shapes along the edges $a$ and $b$.\\
Under pure dilatation, $\sigma_{x}$ and $\sigma_{y}$ can be written in closed form as
\eqb{lll}
\sigma_{x}=\sigma_{y}=(\ds 2n_2+6n_3\,\sJ_{1}+12n_4\,\sJ_{1}^2+20n_5\,\sJ_{1}^3)/\det \bF~.
\eqe
Based on this expression, the variation of the frequencies for a square sheet with $a = b = 250$ nm follows as is shown in Fig.~\ref{f:BPh_Rect_simply_freqeuncy_variation}a.
Under uniaxial stretch along the armchair and zigzag directions, $\sigma_x$ and $\sigma_y$ can be computed from \eqref{e:BPh_sig}\footnote{For uniaxial stretch, $\sigma_{x}=\sigma_{11}$ and $\sigma_{y}=\sigma_{22}$, where $\sigma_{11}$ and $\sigma_{22}$ are the Cartesian components of $\bsig$, which are computed from \eqref{e:BPh_sig}.}. Based on this, the variation of the frequencies against $\lambda_{1}$ follows as is shown for a square sheet in Fig.~\ref{f:BPh_Rect_simply_freqeuncy_variation}b. The material behavior is anisotropic for uniaxial stretch and the frequencies are increasing faster if the sheet is stretched along the zigzag direction instead of the armchair direction. The frequencies increase monotonically up to the instability point, beyond which the material is unstable.
\section{Conclusions}\label{s:BPh_conclusion}
A new hyperelastic material model is proposed for blue phosphorus ($\beta$-P). The model is fully nonlinear and captures anisotropic behaviour of the material. It is written based on a set of invariants that are obtained from the symmetry group of the lattice. The continuum model is calibrated with DFT data. The model is in good agreement with further data sets from DFT results. The model is implemented in the curvilinear finite element membrane formulation of \citet{Sauer2014_01} and thus used for the simulation of nano-indentation. To the best of our knowledge, there is no atomistic potential for blue phosphorus atoms and the proposed model is the only way to simulate specimens at micro-scale. The proposed material model can be extended to finite temperatures based on the new anisotropic thermoelastic shell formulation of \citet{Ghaffari2019_01}.
\section*{Authors' contributions}
RG and RAS derived the continuum material model and designed the continuum examples. RG implemented the finite element formulation, ran the numerical continuum examples, produced all continuum results, and designed the quantum experiments. FS conducted all density functional theory simulations. RAS and MH supervised and directed the research together. The paper has been written by all authors.
\section*{Competing Interests}{The authors have no conflict of interests.}
\section*{Funding}{Financial support from the German Research Foundation (DFG) through grant GSC 111 is
gratefully acknowledged.}
\section*{Acknowledgement}{The authors would like to thank Mr.~Mohammad Sarkari Khorrami for helpful discussions.}
\appendix
\setcounter{figure}{0}
\setcounter{table}{0}
\section{DFT simulations}\label{s:BPh_DFT_simulations}
Monolayer blue phosphorus is a two-dimensional material with hexagonal structure and lattice parameter of 3.28 \AA. The thickness of the monolayer, due to out-of-plane distance of neighbouring phosphor atoms, is 1.23 \AA. The atomic structure of blue phosphorus is shown in Fig.~\ref{f:BPh_atomic_Structure} and Tab.~\ref{t:PhBlu_lattice_str}.
DFT calculations of the mechanical properties of blue phosphorus are conducted using the {\sc Quantum ESPRESSO} package \citep{Giannozzi2009_01,Giannozzi2017_02}. An ultrasoft pseudopotential is used to approximate the effect of non-valence electrons and the exchange correlation energy is approximated using the Perdew, Burke, and Ernzerhof (PBE) exchange-correlation functional \citep{Perdew1996_01}.
The periodic unit cell contains 2 phosphor atoms and -- in order to eliminate the interaction with replicates in $z$ direction -- an interlayer spacing of 25 \AA~is chosen. The Brillouin zone integration is performed within a Monkhorst-Pack method with a $15 \times 15 \times 1$ k-mesh. The kinetic energy cutoff is set to 45 Ry and 360 Ry for wavefunctions and charge density, respectively. The convergence threshold for self-consistency is set to $10^{-8}$ Ry. Geometries have been optimized by the BFGS algorithm until the forces on each atom are less than $10^{-5}$ $\text{R}_{\text{y}}/\text{r}_{\text{Bohr}}$.
\begin{table}[h]
	\centering
	\caption{Lattice structure of blue phosphorus.\label{t:PhBlu_lattice_str}}
	\begin{tabular}{ccc}
		
		\hline \\[-3mm]
		
		Lattice parameter (\AA)  & Thickness (\AA) & Bond length (\AA)\\ [2mm]
		\hline
		$3.28$ & $1.23$ & $2.26$\\[1mm]
		\hline
	\end{tabular}
\end{table}
\section{Derivative of the invariants with respect to logarithmic strain}\label{s:derivative_log_inv}
The first and second derivatives of the invariants $\sJ_{i}$ with respect to the logarithmic strain $\bE^{(0)}$ are needed in the derivation of the stress and elasticity tensors.
The first derivatives are \citep{Kumar2014_01}
\eqb{lll}
\ds \pa{\sJ_{1}}{\bE^{(0)}} \is \bI~,
\eqe
\eqb{lll}
\ds \pa{\sJ_{2}}{\bE^{(0)}} \is \bE^{(0)}_{\text{dev}}~,
\eqe
\eqb{lll}
\ds \pa{\sJ_{3}}{\bE^{(0)}} \is
\ds \frac{3}{8}\biggl\{ \left[\left(\widehat{\bM}:\bE^{(0)}_{\text{dev}}\right)^2 - \left(\widehat{\bN}:\bE^{(0)}_{\text{dev}}\right)^2 \right]\,\widehat{\bM}-2\left(\widehat{\bM}:\bE^{(0)}_{\text{dev}}\right)\left(\widehat{\bN}:\bE^{(0)}_{\text{dev}}\right)\,\widehat{\bN}
\biggr\}
\eqe
while the second derivatives are
\eqb{lll}
\ds \paqq{\sJ_{1}}{\bE^{(0)}}{\bE^{(0)}} \is \sO~,
\eqe
\eqb{lll}
\ds \paqq{\sJ_{2}}{\bE^{(0)}}{\bE^{(0)}} \is \ds \sI-\frac{1}{2}\,\bI\otimes\bI~,
\eqe
\eqb{lll}
\ds \paqq{\sJ_{3}}{\bE^{(0)}}{\bE^{(0)}}\is \ds \frac{3}{4}
\biggl\{
\left(\widehat{\bM}:\bE^{(0)}_{\text{dev}}\right)\left(\widehat{\bM}\otimes\widehat{\bM}\right)-
\left(\widehat{\bN}:\bE^{(0)}_{\text{dev}}\right)\left[ \widehat{\bM}\otimes\widehat{\bN}+\widehat{\bN}\otimes\widehat{\bM}\right] \\[3mm]
\mi \left(\widehat{\bM}:\bE^{(0)}_{\text{dev}}\right)\left(\widehat{\bN}\otimes\widehat{\bN}\right)
\biggr\}~,
\eqe
where $\sO$ is the fourth order zero tensor and $\sI$ is the fourth order identity tensor (see \citet{Kumar2014_01} for $\sO$ and $\sI$).
\section{Finite difference computation of the stress and elasticity tensor}\label{s:Finite_difference_stress_elasticity}
The contra-variant components of the Kirchhoff surface stress tensor $\tauab$ and the corresponding elasticity tensor $c^{\alpha\beta\gamma\delta}$ are needed for the curvilinear membrane FE formulation of \citet{Sauer2014_01}. Here a central difference scheme is proposed for their calculation. $W$ is a function of the logarithmic surface strain tensor $\bE^{(0)}$. $\bE^{(0)}$ can be written as $\bE^{(0)}=1/2\ln\bC$, where $\bC=\auab\,\bA^{\alpha}\otimes\bA^{\beta}$. Thus $W(\auab)$ and accordingly, $\tauab$ can be computed as
\eqb{lll}
\ds \tauab \dis \ds 2\pa{W}{\auab}\approx\frac{W\left(\left[\auab\right]+[\Delta a_{\alpha\beta}^{+}]\right)-W\left(\left[\auab\right]+[\Delta a_{\alpha\beta}^{-}]\right)}{\Delta h}~,
\eqe
where $\left[\auab\right]$ is the matrix form of the current surface metric and $[\Delta a_{\alpha\beta}^{+}]$ are
\eqb{lllllll}
\ds [\Delta a_{11}^{+}] \dis \ds \left[\begin{array}{cc}
                                             \Delta h  & 0\\
                                             \ds 0 & 0 \\
                                           \end{array}\right];~~
\ds [\Delta a_{22}^{+}] \dis \ds \left[\begin{array}{cc}
                                             \ds 0 & \ds 0 \\
                                             \ds 0 & \Delta h \\
                                           \end{array}\right];~~
\ds [\Delta a_{12}^{+}]=[\Delta a_{21}^{+}] \dis \ds \left[\begin{array}{cc}
                                             0 & \ds \frac{1}{2}\Delta h \\
                                             \ds \frac{1}{2}\Delta h & 0 \\
                                           \end{array}\right]~,
\eqe
and $[\Delta a_{\alpha\beta}^{-}]=-[\Delta a_{\alpha\beta}^{+}]$. The corresponding elasticity tensor is
\eqb{lll}
\ds c^{\alpha\beta\gamma\delta} \dis \ds 4\paqq{W}{\auab}{\augd} \approx \frac{\tauab\left(\Delta\augd^{+}\right)-\tauab\left(\Delta\augd^{-}\right)}{\Delta h}~,
\eqe
where $\tauab(\Delta\augd^{+})$ and $\tauab(\Delta\augd^{-})$ are defined by
\eqb{lll}
\ds \tauab(\Delta\augd^{+}) \dis \ds \frac{W\left(\left[\auab\right]+[\Delta a_{\alpha\beta}^{+}]+[\Delta a_{\gamma\delta}^{+}]\right)- W\left(\left[\auab\right]+[\Delta a_{\alpha\beta}^{-}]+[\Delta a_{\gamma\delta}^{+}]\right)}{\Delta h}
\eqe
and
\eqb{lll}
\ds \tauab(\Delta\augd^{-}) \dis \ds \frac{W\left(\left[\auab\right]+[\Delta a_{\alpha\beta}^{+}]+[\Delta a_{\gamma\delta}^{-}]\right)- W\left([\auab]+[\Delta a_{\alpha\beta}^{-}]+[\Delta a_{\gamma\delta}^{-}]\right)}{\Delta h}~.
\eqe
In these formulae, $\auab$ and $\Delta \auab$ can be replaced by the Cartesian components of $\bC$ and its increment $\Delta \bC$, in order to compute the Cartesian components of the second Piola-Kirchhoff surface stress tensor $\bS$ and its conjugate elasticity tensor $\bbC$ since $\bS=2\partial W /\partial \bC$ and $\bbC=\ds 4\partial^2 W/\partial\bC\partial\bC$. In the computation of $\tauab(\Delta\augd^{+})$, the components of $[\auab]+[\Delta a_{\alpha\beta}^{+}]+[\Delta a_{\gamma\delta}^{+}]$ are summed. For example as
\eqb{lll}
\ds \tau^{12}\left(\Delta a_{21}^{+}\right) \dis \ds \frac{W\left([a_{12}]+[\Delta a_{12}^{+}]+[\Delta a_{21}^{+}]\right)- W\left([a_{12}]+[\Delta a_{12}^{-}]+[\Delta a_{21}^{+}]\right)}{\Delta h}~.
\eqe
\bibliographystyle{model1-num-names}
\bibliography{bibliography}
\end{document}